\documentclass[12pt]{article} 
\usepackage[utf8]{inputenc}
\usepackage{mathtools}
\usepackage{amssymb}
\usepackage{amsthm}
\usepackage{graphicx}
\usepackage{verbatim}
\usepackage{hyperref}
\usepackage{pgfplots}
\usepackage{subfigure}
\usetikzlibrary{calc}
\usepgfplotslibrary{fillbetween}
\usepackage[inline]{enumitem}
\usepackage{geometry,caption,color,setspace,comment,footmisc,pdflscape,subfigure,array}
\pgfplotsset{width=0.9\textwidth,compat=1.8}

\usepackage[
backend=biber, natbib,
style=authoryear-comp,
maxcitenames=2
]{biblatex}
\DeclareSourcemap{
  \maps[datatype=bibtex, overwrite]{
    \map{
      \step[fieldset=urldate, null]
      \step[fieldset=note, null]
      \step[fieldset=abstract, null]
      \step[fieldset=url, null]
      \step[fieldset=issn, null]
      \step[fieldset=eprint, null]
      \step[fieldset=doi, null]
      \step[fieldset=keywords, null]
      \step[fieldset=number, null]
    }
  }
}

\theoremstyle{plain}
\newtheorem{theorem}{Theorem}
\newtheorem{lemma}{Lemma}
\newtheorem{corollary}{Corollary}
\newtheorem{prop}{Proposition}
\newtheorem{assumption}{Assumption}

\theoremstyle{definition}
\newtheorem{definition}{Definition}
\newtheorem{examplex}{Example}

\newenvironment{example}
  {\pushQED{\qed}\examplex}
  {\popQED\endexamplex}

\begin{document}

\begin{titlepage}
\title{Asymmetric All-Pay Auctions with Spillovers\footnote{We thank Wojciech Olszewski, Alessandro Pavan, Marciano Siniscalchi, Bruno Strulovici, and Asher Wolinsky for invaluable comments throughout the writing process.}}
\author{Maria Betto\footnote{Maria Betto: Northwestern University, maria.betto@u.northwestern.edu, \href{https://mariabetto.com}{mariabetto.com}} \and Matthew W. Thomas\footnote{Matthew W. Thomas: Northwestern University, mthomas@u.northwestern.edu, \href{https://mwt.me}{mwt.me}}}
\date{\today}
\maketitle
\begin{abstract}
\noindent When opposing parties compete for a  prize, the sunk effort players exert during the conflict can affect the value of the winner's reward. These \textit{spillovers} can have substantial influence on the equilibrium behavior of participants in applications such as lobbying, warfare, labor tournaments, marketing, and R\&D races. To understand this influence, we study a general class of asymmetric, two-player all-pay auctions where we allow for spillovers in each player's reward.  The link between participants' efforts and rewards yields novel effects -- in particular, players with higher costs and lower values than their opponent sometimes extract larger payoffs.\\

\vspace{0in}

\bigskip
\end{abstract}
\setcounter{page}{0}
\thispagestyle{empty}
\end{titlepage}
\pagebreak \newpage

\section{Introduction}

All-pay auctions, or contests, model strategic interactions among players who must expend some non-refundable effort in order to win a prize. They have been applied in diverse settings such as labor \parencite{rosen_prizes_1986}, R\&D races \parencite{dasgupta1986theory,che1998caps}, and litigation \parencite{baye2005comparative}. For tractability, the recent literature mostly assumes that players' actions affect their opponent's probability of winning, but not the value of the prize. Yet, in many settings, such \textit{spillover effects} on the prizes themselves arise naturally.

For example, consider the setting in \citet{che1998caps}, where two lobbyists compete in an all-pay auction to win an incumbent politician's favor through campaign contributions. If the politician were instead a candidate running for office, then she would only be able to provide the reward if successfully elected. In this case, it is natural to assume that total campaign contributions increase the candidate's chances of prevailing. Therefore, each lobbyist's contributions increase her opponent's value for winning the politician's political favor. This raises new questions: is it better to curb one's own contributions to make their opponent lose interest? Or is it preferable to ramp up the competition? These questions have been largely left unanswered.

In other settings, spillovers may be designed. Consider an all-pay version of a standard labor tournament, in which division managers apply effort towards some production technology in order to win a promotion awarded to the most productive division. To maximize aggregate effort, a principal might choose to make the value of this promotion depend on everyone's performance in the contest. For example, if the promotion is for a partnership or involves stock options, the prize will be increasing in the efforts of all players. The effect that such compensation schemes have on the equilibrium has not yet been studied.

This paper fully identifies the equilibrium strategies and payoffs in general two-player auctions with spillovers and establishes their uniqueness.\footnote{This paper also establishes the existence of equilibrium, though this result has already been proven; see \citet{olszewski2022equilibrium}, for example. Our method, however, differs substantially from the previous literature.} We consider games with (i) complete information, (ii) deterministic prizes, (iii) at least partially sunk investment costs, and (iv) a general dependence of each participant's value for the prize on both players' actions. The key contribution of this paper lies in incorporating (iv). Indeed, all-pay contests without spillovers were extensively studied by \citet{siegel2009all,siegel2010asymmetric}. These papers fully characterize equilibrium strategies and payoffs in games where contestants incur some (partially) unrecoverable cost, such as effort, in order to compete for prizes. We generalize the two-player, single-prize version of their model to allow for general spillovers to affect the winner's payoff. Our paper also has some  overlap with the symmetric linear contests with spillovers studied in \citet{baye2012contests}. Unlike their work, however, we restrict attention to the all-pay case, but allow for asymmetric equilibria and nonlinear payoffs. Even in the symmetric, linear all-pay auction with spillovers, we note that no previous paper that we are aware of has established equilibrium uniqueness.

The addition of spillovers can have a significant impact on equilibrium behavior. First, players with strictly higher costs can have higher payoffs than those with lower costs, even if their value functions for the prize are identical. In fact, in some settings, players could increase their payoffs if they were allowed to commit to a schedule of costly handicaps (See Section \ref{sec:payoffs}). Thus, trying to favor an ``underdog'' participant in a contest by means of reducing their costs may very well have the opposite effect, and in fact decrease their welfare in equilibrium. This is also important in settings in which players can commit to increasing their costs (e.g. by selecting an inefficient technology), as they may choose to do so.

Another contribution of this paper is the procedure to construct equilibrium strategy profiles. The equilibrium strategy distributions of asymmetric all-pay contests have two distinct parts: the densities and a mass point at zero. In the literature on all-pay contests without spillovers, starting with \citet{baye1996all}, expected payoffs are obtained independently of the equilibrium distribution. This independence is exploited to derive the probability mass at zero for the weaker player from the payoffs, which is then used to compute the densities. In the presence of spillovers, however, a player's payoffs cannot be derived without the equilibrium strategy of their opponent. Because of this, the same process cannot be followed. To overcome this difficulty, we introduce an algorithm that works in exactly the opposite order: first, it solves for the density independently of the mass point, and then uses this density to find the probability mass at zero.

Our method capitalizes on the theory of Volterra Integral Equations (VIEs), which are integral equations with a unique fixed-point that can be obtained via iteration. To the best of our knowledge, these techniques have not previously been applied to the determination of equilibrium mixed-strategy profiles.\footnote{Few other works in Economics use VIE methods in general. We note \citet{mcafee1989extracting} and \citet{mcafee1992correlated} as some early examples. More recently, \citet{GOMES2014421} also used VIEs, to compute the unique efficient equilibrium bidding functions in generalized second-price auctions.}

The game we study is general enough to encompass many different applications in which spillovers matter. In particular, investment wars, contests with winner's regret, and militaristic conflicts all fit our framework, since spillovers are key in each of these settings. Our model also subsumes a natural extension to the war of attrition which, unlike the classical model, yields a unique equilibrium on a bounded support. We are also able to use the same framework to describe wars of attrition where rational agents face uncompromising (never-yielding) types with positive probability, as in \citet{abreu2000bargaining} and \citet{kambe2019n}. Our approach identifies why these games admit unique equilibria when the regular war of attrition does not: the addition of an uncompromising type introduces an unavoidable cost that depends on a player's own score, and we show this single characteristic is sufficient in ensuring a unique equilibrium.

Finally, we extend the analysis to more than two players. The uniqueness result does not hold when the number of bidders exceeds two. We are nonetheless able to characterize a class of asymmetric equilibria when (appropriately normalized) costs are ranked. In this case, we show only two players participate in equilibrium. In addition, we are able to fully characterize the unique symmetric equilibrium for all-pay auctions with (i) more than two identical players, (ii) multiple homogeneous prizes, and (iii) spillovers generated by the first runner-up. This setting accommodates a broad class of games including wars of attrition and auctions with winner's regret with any number of players and prizes. This extends the usefulness of our novel methodology.

The paper is organized as follows. We introduce the model, the equilibrium concept and the assumptions in Section \ref{sec:model}. We construct the equilibrium and prove its uniqueness in Section \ref{sec:solution}. Section \ref{sec:payoffs} presents sufficient conditions under which a player has a positive expected payoff. This includes an example where a player with higher costs and lower values receives a positive expected payoff, while her opponent receives zero. Section \ref{sec:closed-forms} contains useful results on closed forms that allows for simplified equilibrium computations in certain special cases. We illustrate their usage in the following Section \ref{sec:other-applications}, which is dedicated to applications. Sections \ref{sec:off_def_balance}, \ref{sec:winners_regret} and \ref{sec:war_of_inv} in particular showcase closed-form solutions. Section \ref{sec:WoAcp} introduces a general perturbation of the classic war of attrition that ensures the equilibrium is unique. This perturbation admits the war of attrition with the possibility of an uncompromising type as a special case. In Section \ref{sec:more_players}, we extend the analysis to contests with more than two players. Uniqueness no longer holds generally, though we are still able to find the unique symmetric equilibrium of a $n$-player, $m$-prize all-pay auction with spillovers. Finally, in Section \ref{sec:conclusion}, we review the related literature and discuss the results.

\section{Model}\label{sec:model}

We focus, for now, on auctions with two participants. Extensions with more players are considered in Section \ref{sec:more_players}, where we show that the symmetric equilibrium of an auction with any number of identical players and prizes is just a transformation of the equilibrium of the two-player case.

An asymmetric auction with spillovers is a family $\{I,\{\tilde{S}_i\}_{i\in I}, \{u_{i}\}_{i\in I}\}$, where
\begin{enumerate}
    \item $I:=\{1,2\}$ is the index set of players.
    \item For each $i\in I$, $\tilde{S}_i:=[0,\infty)$ is Player $i$'s action space, i.e. her set of available scores (or bids). We use a tilde because a later assumption will allow us to replace the action set with a bounded interval. We let $s_{-i}$,$\tilde{S}_{-i}$ denote the action and action space, respectively, of Player $j \neq i$.
    \item For each $i\in I$, $u_i:\tilde{S}\to \mathbb{R}$ is Player $i$'s payoff, where $\tilde{S}:=\prod_{i\in I}\tilde{S}_i$. 
    
    Let $s:=(s_i;s_{-i})$ denote an arbitrary element of $\tilde{S}$. Then, for each $(s_i;s_{-i})$, we further define
    \[
    u_i (s_i;s_{-i}) := p_i(s_i;  s_{-i}) v_i( s_i;  s_{-i} ) - c_i(s_i)
    \]
     where (i) $p_i(s_i;  s_{-i})$ denotes the probability that $i$ wins the prize given the score profile $(s_i;  s_{-i})$, with $p_i(s_i;  s_{-i}) = 1 - p_{-i}(s_{-i};  s_{i})$ and
    \begin{align*}
         p_i(s_i;  s_{-i}) = \left\{\begin{array}{ll}
            1 & \text{ if } s_{i} >  s_{-i}, \\
            \alpha_i \in [0,1] & \text{ if } s_{i} =  s_{-i}, \\
            0 & \text{ if } s_{i} <  s_{-i};
        \end{array}\right.
        \end{align*}
    (ii)  $v_i:\tilde{S}\to\mathbb{R}_{+}$ maps each score profile $(s_i;s_{-i})$  to Player $i$'s value $v_i( s_i;  s_{-i} )$ from winning the prize, and (iii) $c_i:\tilde{S}_i\to\mathbb{R}_+$ outputs Player $i$'s private cost $c_i(s_i)$ given her submitted score $s_i$.
\end{enumerate}

\begin{definition}[Two-player all-pay auction with spillovers]
    A two-player all-pay auction is said to have \emph{spillovers} if, for some $i\in I$ and $s_i\in\tilde  S_{i}$, there exists $s_{-i},\hat{s}_{-i}\in\tilde{S}_{-i}$  such that
    \[
        v_i(s_i,s_{-i})\neq v_i(s_i,\hat{s}_{-i})
    \]
    i.e., the prize's value for at least one player and an action of that player is not constant in their opponent's action.
\end{definition}

Accommodating spillovers is the distinguishing feature of our analysis. As is standard, we are interested in characterizing the Nash equilibrium of these general contests.

\begin{definition}[Best-responses] 
    Consider a two-player all-pay auction $\{I,\{\tilde{S}_i\}_{i\in I}, \{u_{i}\}_{i\in I}\}$. For each $i\in I$, let $\Delta \tilde{S}_i$ denote the set of probability distributions on $\tilde{S}_i$ and let $\Delta \tilde{S} := \prod_{i\in I} \Delta \tilde{S}_i$. Player $i$'s best response set $b_i({G}_{-i})$ to ${G}_{-i}\in \Delta \tilde{S}_{-i} $ is given by
    \[
        b_i({G}_{-i}):=\arg\max_{s\in \tilde{S}_i}\int_{\tilde{S}_{-i}} u_i(s;s_{-i})d{G}_{-i}(s_{-i})
    \]
\end{definition}

\begin{definition}[Nash equilibrium] \label{def:eq} 
    Consider the two-player all-pay auction $\{I,\{\tilde{S}_i\}_{i\in I}, \{u_{i}\}_{i\in I}\}$. A Nash equilibrium of this game is a profile $\mathbf{G}^\star:=(G_i^\star)_{i\in I}\in \Pi_{i \in I} (\Delta \tilde{S}_i)$ where, for each $i\in I$, $G_i^\star$'s induced probability measure assigns measure one to $b_{i}({G}_{-i}^\star)$.
\end{definition}

\subsection{Assumptions}\label{sec:assumptions}

The following assumptions are imposed throughout whenever a two-player all-pay auction is invoked. \ref{appendix:assumptions} shows that none of these assumptions are superfluous to our results.

\begin{assumption}[A\ref{cond:smo}, Smoothness]
\label{cond:smo} The function $v_i(s_i; {y})$ is continuously differentiable in $s_i$ and continuous in ${y}$ for all $i\in I$, $s_i \in \tilde{S}_i$, and ${y}\in \tilde{S}_{-i}$ with $s_i \geq y$. The function  $c_i(s_i)$ is continuously differentiable in $s_i$ for all $i\in I$, $s_i \in \tilde{S}_i$.
\end{assumption}

\begin{assumption}[A\ref{cond:mon}, Monotonicity]
\label{cond:mon} For all $i\in I$ and $s_i > 0$, $c'_i(s_i) > 0$ and 
\[
    v'_i(s_i; y) < c'_i(s_i)
\]
for almost all $y$, where $v'_i(s; {y}) := \partial v_i(s_i; {y})/\partial s_i$.
\end{assumption}

\begin{assumption}[A\ref{cond:int}, Interiority]
\label{cond:int}  For all $i\in I$, 
\[
   v_i(0, 0) > c_i(0) = 0
   \quad \text{ and } \quad
   \lim_{s_i \to \infty} \sup_{y\in \tilde{S}_{-i}} v_i(s_i; {y}) < \lim_{s_i \to \infty} c_i(s_i).
\]
\end{assumption}

Versions of assumptions A\ref{cond:smo}, A\ref{cond:mon}, and A\ref{cond:int} are adopted by most papers in the all-pay auction literature. A\ref{cond:mon} formalizes the sense in which these contests are all-pay, since bids are costly for both the winner and the loser.\footnote{We note that A\ref{cond:mon} does exclude situations where a higher score is not necessarily more costly. \citet{siegel2014contests} discusses contests with nonmonotonic costs, allowing for competitors with head starts and the provision performance-based subsidies. These contingencies are excluded from our analysis.} A\ref{cond:int} ensures that bids are positive and bounded.

Note that, for each $i\in I$, there exist $T_i \in \tilde{S}_i$ such that Player $i$ will never choose a score $s \geq T_i$. Thus, we can restrict the action space to $S_i:=[0,T_i]$.

\begin{assumption}[A\ref{cond:pos}, Discontinuity at ties] \label{cond:pos} For all $i\in I$ and $s\in {S}_{i}\cap{S}_{-i}$,
\[
   v_i(s; s) > 0.
\]
\end{assumption}

Assumption A\ref{cond:pos} is a novel, yet natural assumption. It states that agents would prefer to win a tie than lose one. It is satisfied if the prize is always valuable (i.e. winning is better than losing), or if there are no spillovers. To see that it is never violated in the absence of spillovers, note that $T_i$ is less than or equal to any $x$ satisfying $v_i(x; y) \leq c_i(x)$ for all $y\leq x$. If there are no spillovers and $v_i(s) \leq 0$ for some $s \leq T_i$, then $c_i(s) \leq 0$. Therefore $s = 0$, which violates Assumption A\ref{cond:int}. Note that this assumption is equivalent to assuming a discontinuity in payoffs at ties because A\ref{cond:smo} and A\ref{cond:int} guarantee that $v_i(s; s) \neq 0$ implies A\ref{cond:pos}.

\section{Characterization of equilibrium}\label{sec:solution}

By standard arguments contained in the Appendix, any pair of equilibrium strategies will be mixed with support on some interval $[0, \bar{s}]$, and at most one player will have a mass point at zero. Players must therefore be indifferent between all points of their interval support:
\begin{equation}\label{eq:ind}
    \bar{u}_i(G_{-i}) := \int_{0}^{s}v_i(s;y) \, dG_{-i}(y) - c_i(s) \qquad \text{ for all }s\in[0,\bar{s}].
\end{equation}

Any pair of distributions ($G_1, G_2$) that satisfy \eqref{eq:ind} is an equilibrium. This paper's main contribution to the literature is in characterizing the solution to this system of equations, and in showing that it is unique.

\begin{theorem} \label{thm:main}
    Every two-player all-pay auction has a \textbf{unique Nash equilibrium} $(G^\star_i)_{i\in I} \in \prod_{i\in I}(\Delta S_i)$ in mixed strategies. Furthermore, 
    \begin{equation} \label{eq:eqstr}
        G^\star_i(s) = \int_0^{s} \tilde{g}_i(y) \, dy + \int_{\bar{s}}^{\bar{s}_i} \tilde{g}_i(y) \, dy,
    \end{equation}
    where $\tilde{g}_{i}(s)$ solves
    \begin{equation}\label{eq:ieq}
        \tilde{g}_{i}(s) = \frac{c'_{-i}(s)}{v_{-i}(s; s)} - \int_0^s \frac{v'_{-i}(s; y)}{v_{-i}(s; s)} \tilde{g}_{i}(y) \, dy,
    \end{equation}
    $\bar{s}_i$ solves $\int_0^{\bar{s}_i} \tilde{g}_i(y) dy = 1$ and $\bar{s} = \min_{i\in I} \bar{s}_i$. The solution admits the following representation
    \[
        \tilde{g}_{i}(s) = \frac{c'_{-i}(s)}{v_{-i}(s; s)} + \int_0^{s} r_{-i}(s;y) \frac{c'_{-i}(y)}{v_{-i}(y; y)} \, dy,
    \]
    where
    \[
        r_{-i}(s;y) := - k_{-i}^0(s;y) + k_{-i}^1(s,y) - k_{-i}^2(s;y) + \dots
    \]
    for $k_{-i}^0(s;y) := \frac{v_{-i}'(s;y)}{v_{-i}(s;s)}$ and $k_{-i}^n(s;y)$, $n=1,2,\dots$, defined recursively by
    \[
    k_{-i}^n(s;y) := \int_y^s{\frac{v_{-i}'(s;z)}{v_{-i}(s;s)}k_{-i}^{n-1}(z;y)}\, dz.
    \]
\end{theorem}
    
We outline the proof here with an emphasis on the general methodology. We show in the appendix that in any equilibrium, players choose strictly increasing, continuous mixed strategies with common support on some interval $[0, \bar{s}]$, as in \eqref{eq:ind}, and that at most one participant can have a mass point at zero. Moreover, differentiating \eqref{eq:ind} yields \eqref{eq:ieq}, which must be satisfied on $[0, \bar{s}]$ in equilibrium for some $\bar{s}$ (Lemma \ref{prop:interval} in the Appendix).

The key step is recognizing that we can apply results about Volterra Integral Equations (VIE) to show that  \eqref{eq:ieq} has a unique solution. The relevant result is summarized in the following Lemma. For a proof, see e.g. \citet{brunner2017volterra}.

\begin{lemma}[\citet{volterra1896sulla}]\label{prop:invert}
    Let $K(s; y)$ and $f(s)$ be continuous functions. Then, the following integral equation
    \begin{equation}\label{eq:vie}
        g(s) = f(s) + \int_0^s K(s;y) g(y) dy \qquad \text{ for all }s\in[0,\bar{s}]
    \end{equation}
    has a solution, $g$, unique almost everywhere. Moreover, \eqref{eq:vie} defines a contraction mapping, implying the solution can be found by iteration. This iteration reduces to:
    \[
        g(s) = f(s) + \int_0^s R(s; y) f(y) dy,
    \]
    where $R(s; y)$ is the unique resolvent kernel defined by
    \[
        R(s; y) = \sum_{m=0}^\infty K_m(s; y)
    \]
    where $K_0 \equiv K$ and $K_m$ is defined recursively for $m=1,2,\dots$ as
    \[
        K_m(s;y) = \int_y^s K_{m-1}(s; z) K(z; y) \,dz.
    \]
\end{lemma}

Note that \eqref{eq:vie} is the same as \eqref{eq:ieq} for $f(s) := c'_{-i}(s)/v_{-i}(s; s)$ and $K(s;y) := - v'_{-i}(s; y)/v_{-i}(s; s)$. So, Lemma \ref{prop:invert} implies that only one pair of functions $(\tilde{g}_1, \tilde{g}_2 )$ solves \eqref{eq:ieq}. Next we show that the unique solutions are densities, i.e. for each $i$ there is an interval $[0, \bar{s}_i]$ where $\tilde{g}_i$ is non-negative and integrates to one.

\begin{lemma}\label{prop:pdf}
    Assume a two-player all-pay auction where $(\tilde{g}_i)_{i\in I}$ satisfies the indifference condition in \eqref{eq:ieq}. Then, for each $i\in I$, there exists $\bar{s}_i\in S_i$ such that 
    \begin{equation}\label{eq:sbari}
        \int_0^{\bar{s}_i} \tilde{g}_i(y) dy = \tilde{G}_i(\bar{s}_i) = 1,
    \end{equation}
    and $\tilde{g}_i(s)$ is positive for $s \leq \bar{s}_i$.
\end{lemma}

Lemma \ref{prop:pdf} is proven in the Appendix. We must now ensure the two densities have the same support. The next key insight is that there is exactly one way to do this. Recall that at most one player can have an mass point and that this mass point must be at zero (Lemma \ref{prop:interval}). If $\bar{s}_1 = \bar{s}_2$, then there is a unique equilibrium without any mass point. Otherwise, order the players such that $\bar{s}_1 < \bar{s}_2$. Then, give Player 2 a mass point of size $1 - \tilde{G}_2(\bar{s}_1)$. By construction, both players' densities integrate to one on the common support $[0, \bar{s}_1]$.

The above can be performed via the following steps: 
\begin{enumerate}
    \item Find each $\tilde{g}_i(s)$.\footnote{Analytically, it can be expressed as a series or in closed form when possible -- see Section \ref{sec:closed-forms} -- or numerically -- see Appendix \ref{appendixnum}.}
    \item Integrate each $\tilde{g}_i(s)$ to find $\bar{s}_i$ given by equation \ref{eq:sbari}.
    \item Take $\bar{s} = \min_i \bar{s}_i$ and give each player an mass point at zero of size
    \[
        1 - \tilde{G}_i(\bar{s}),
    \]
    which is positive for at most one player.
\end{enumerate}

\begin{figure}
\centering
\subfigure[Step 1]
    {
        \begin{tikzpicture}
        \begin{axis}[title={$\tilde{g}_1(s)$}, 
        name=g1, 
        height = 5.5cm, 
        width = 8.5cm, 
        xlabel = Score, 
        xmin=0, 
        xmax = 3, 
        ymin=0, 
        ymax = 4,
        restrict y to domain=-5:15,
        samples = 100, 
        ticks = none, 
        mark = none,
        ]
            \addplot[smooth, black] {(2/0.5)*((x*(0.5+1)/0.5)+1)^((-0.5/(0.5+1))-1)};
        \end{axis}
        \begin{axis}[title={$\tilde{g}_2(s)$},
        name=g2, 
        at={($(g1.east)+(1cm,0)$)},
        anchor=west,
        height = 5.5cm,
        width=8.5cm, 
        xlabel = Score, 
        xmin = 0, 
        xmax = 3,
        ymin=0, 
        ymax = 4,
        restrict y to domain=-5:15,
        samples = 100, 
        ticks = none, 
        mark = none,
        ]
            \addplot[smooth, black] {(1.5/1)*((x*(0.5+1)/1)+1)^((-0.5/(0.5+1))-1)};
        \end{axis}
        \end{tikzpicture}
    }
    \subfigure[Step 2]
    {
        \begin{tikzpicture}
        \begin{axis}[title={$\tilde{g}_1(s)$}, 
        name=g1, 
        height = 5.5cm, 
        width = 8.5cm, 
        xlabel = Score, 
        xmin=0, 
        xmax = 3, 
        ymin=0, 
        ymax = 4,
        restrict y to domain=-5:15,
        samples = 100, 
        ticks = none, 
        mark = none,
        ]
            \addplot[name path=pdf, smooth, color=black] {(2/0.5)*((x*(0.5+1)/0.5)+1)^((-0.5/(0.5+1))-1)};
            \path [name path = xaxis] (axis cs:0,0) -- (axis cs:1,0);
            \addplot[fill = black, fill opacity = 0.12] fill between[of=pdf and xaxis, soft clip ={domain=0:0.457}];
            \addplot[smooth, black] coordinates { (0.457,0) 
            (0.457,4) };
        \end{axis}
        \begin{axis}[title={$\tilde{g}_2(s)$},
        name=g2, 
        at={($(g1.east)+(1cm,0)$)},
        anchor=west,
        height = 5.5cm,
        width=8.5cm, 
        xlabel = Score, 
        xmin = 0, 
        xmax = 3,
        ymin = 0, 
        ymax = 4,
        restrict y to domain=-5:15,
        samples = 100, 
        ticks = none, 
        mark = none,
        ]
            \addplot[name path=pdf2, smooth, black] {(1.5/1)*((x*(0.5+1)/1)+1)^((-0.5/(0.5+1))-1)};
            \path [name path = xaxis] (axis cs:0,0) -- (axis cs:2,0);
            \addplot[fill = black, fill opacity = 0.12] fill between[of=pdf2 and xaxis, soft clip ={domain=0:1.583}];

            \addplot[smooth, black] coordinates { (1.583,0)  (1.583,4) };
        \end{axis}
        \end{tikzpicture}
    }
    \subfigure[Step 3]
    {
        \begin{tikzpicture}
        \begin{axis}[title={$\tilde{g}_1(s)$}, 
        name=g1, 
        height = 5.5cm, 
        width = 8.5cm, 
        xlabel = Score, 
        xmin=0, 
        xmax = 3,
        ymin=0,
        ymax=4,
        restrict y to domain=-5:15,    
        samples = 100, 
        ticks = none, 
        mark = none,
        ]
            \addplot[name path=pdf, smooth, color=black] {(2/0.5)*((x*(0.5+1)/0.5)+1)^((-0.5/(0.5+1))-1)};
            \path [name path = xaxis] (axis cs:0,0) -- (axis cs:1,0);
            \addplot[fill = black, fill opacity = 0.12] fill between[of=pdf and xaxis, soft clip ={domain=0:0.45679012345}];
            \addplot[smooth, black] coordinates { (0.457,0) 
            (0.457,4) };
        \end{axis}
        \begin{axis}[title={$\tilde{g}_2(s)$},
        name=g2, 
        at={($(g1.east)+(1cm,0)$)},
        anchor=west,
        height = 5.5cm,
        width=8.5cm, 
        xlabel = Score, 
        xmin = 0, 
        xmax = 3,
        ymin=0, 
        ymax = 4,
        restrict y to domain=-5:15,
        samples = 100, 
        ticks = none, 
        mark = none,
        ]
            \addplot[name path=pdf2, smooth, black] {(1.5/1)*((x*(0.5+1)/1)+1)^((-0.5/(0.5+1))-1)};
            \path [name path = xaxis] (axis cs:0,0) -- (axis cs:2,0);
            \addplot[fill = black, fill opacity = 0.12] fill between[of=pdf2 and xaxis, soft clip ={domain=0:1.583}];

            \addplot[smooth, black] coordinates { (1.583,0)  (1.58333333333,4) };
            \addplot[smooth, black] coordinates { (0.457,0) 
            (0.45679012345,4) };
            \addplot[fill = black, fill opacity = 0.15] fill between[of=pdf2 and xaxis, soft clip ={domain=0.457:1.583}];
        \end{axis}
        \draw [<->] (9.05,1.5) -- node [above] {mass point} (11.45,1.5);
        \end{tikzpicture}
    }
    \caption{The steps for finding the equilibrium strategies $g_1$ and $g_2$. Begin with $\tilde{g_1}$ and $\tilde{g}_2$ (Step 1); find the cutoffs where each $\tilde{g}_i$ integrates to 1 (Step 2). Finally, enforce identical supports by transferring any excess density to zero (Step 3). This contest is $v_1(s_1; s_2) = 2 + s_1 + 2 s_2$, $v_2(s_2; s_1) = 1 + s_2 + 2 s_1$, $c_1(s_i)= 3 s_1$, and $c_2(s_2) = 4 s_2$.}\label{fig:outline}
\end{figure}
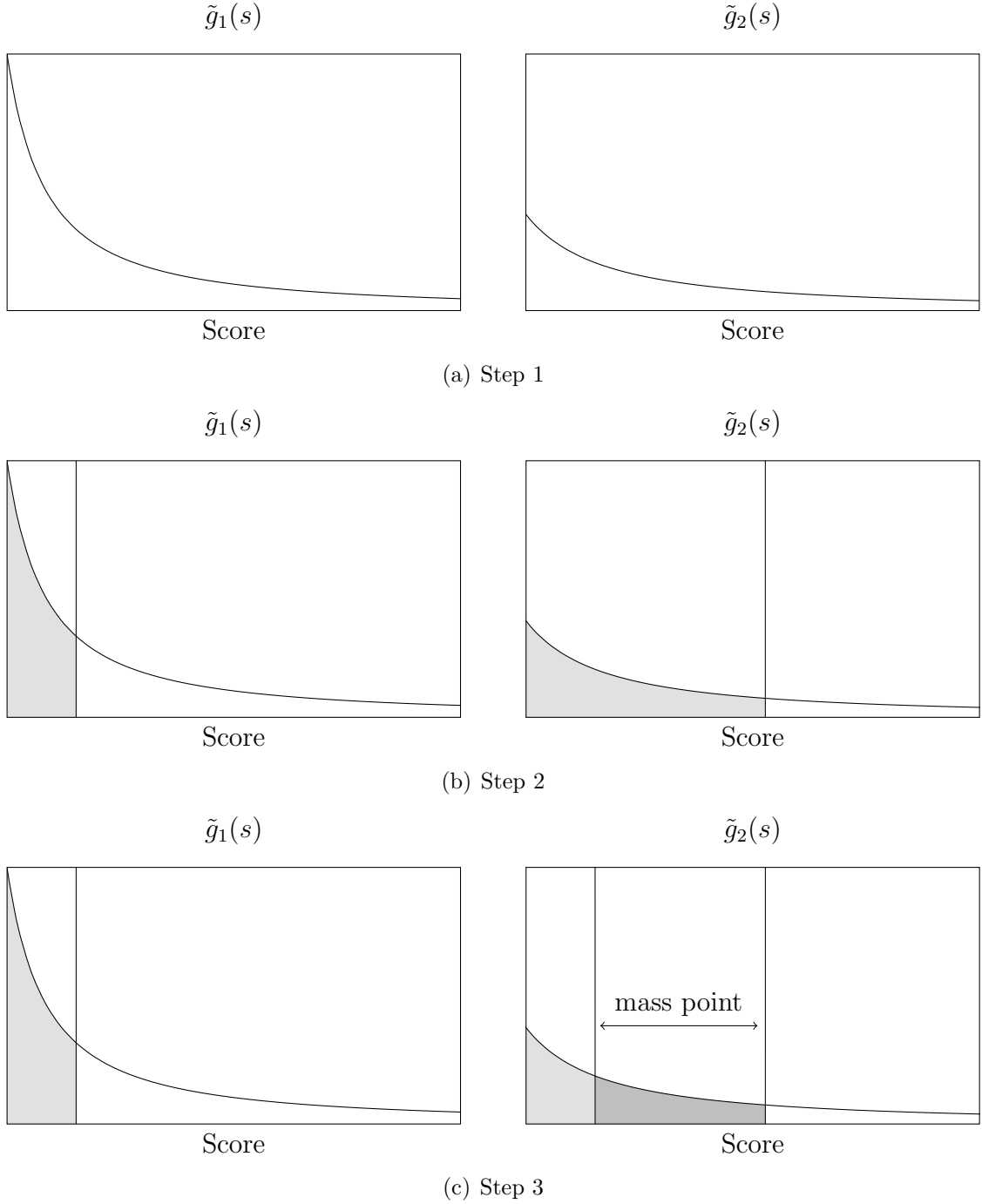

The three steps are illustrated by Figure \ref{fig:outline}.

Since the cumulative distribution functions are useful, we sometimes use the alternate expression presented in Corollary \ref{prop:bigG}.

\begin{corollary}\label{prop:bigG}
    Consider a two-player all-pay auction where $v_i(s; y)$ is continuously differentiable in both arguments for all $i \in I$ (A\ref{cond:smo} guarantees differentiability in the first argument). Then, we can alternatively express the unique equilibrium as
    \[
        G_i(s) = \left[\tilde{G}_i(\overline{s}_i) - \tilde{G}_i(\overline{s})\right] + \tilde{G}_i(s), 
    \]
    where
    \begin{equation}\label{eq:bigG}
        \tilde{G}_{i}(s) =\frac{c_{-i}(s)}{v_{-i}(s; s)} + \int_0^s \frac{\partial v_{-i}(s; y)}{\partial y} \frac{\tilde{G}_{i}(y)}{v_{-i}(s; s)} dy.
    \end{equation}
    The solution admits the following series representation
    \[
        \tilde{G}_{i}(s) = \frac{c_{-i}(s)}{v_{-i}(s; s)} + \int_0^{s} \frac{c_{-i}(y)}{v_{-i}(y; y)} \frac{R_{-i}(s;y)}{v_{-i}(s; s)} \, dy
    \]
    where
    \[
        R_{-i}(s;y) := K_{-i}^0(s;y) + K_{-i}^1(s,y) + K_{-i}^2(s;y) + \dots
    \]
    for $K_{-i}^0(s;y) := \partial v_{-i}(s;y)/\partial y$ and $K_{-i}^n(s;y)$, $n=1,2,\dots$, defined recursively by
    \[
    K_{-i}^n(s;y) := \int_y^s{\frac{\partial v_{-i}'(s;z)}{\partial z }\frac{K_{-i}^{n-1}(z;y)}{v_{-i}(z;z)}}\, dz.
    \]
\end{corollary}

We end this section with a note on parallels between our methodology and the one used in the incomplete information, all-pay auction literature. The similarities are formal in nature, and arise because both problems involve solving a pair of differential (in the incomplete information case) or integral (in our case) equations.

In the incomplete information setting of e.g. \citet{amann1996asymmetric}, the unique equilibrium -- in \textit{pure strategies} -- is obtained as the solution to a pair of differential equations, which arise from taking first-order conditions of each players' expected payoffs. To back out the mass of players types' that bid zero, the authors then make use of the boundary condition where each players' top type must, in equilibrium, choose identical top bids.

In our setting with complete information, the unique equilibrium is instead in \textit{mixed strategies}. It is obtained as the solution to a pair of integral equations, which arise from \textit{indifference}, rather than first-order conditions: players' payoffs must be invariant to any choice of bids within their mixed-strategy supports (Equation \ref{eq:ind}). We are then able to pin down the probability mass with which one of the two players bids zero through a different sort of ``boundary condition''-- specifically, the fact that both players' bidding distributions' supports must be identical, and integrate to 1 in that support.

\section{Payoffs}\label{sec:payoffs}

Since payoffs are constant on the interval $[0,\bar{s}]$, each player $i$ receives an expected payoff of $v_i(0; 0) G_{-i}(0)\geq 0$. Only one player can have a mass point (at zero), so there can be at most one player -- their opponent -- with a positive payoff. In a symmetric contest, both players receive an expected payoff of zero. Theorem \ref{thm:main} immediately implies a necessary and sufficient condition for a player to have a positive payoff.

\begin{corollary}\label{cor:payoff-iff}
    Consider a two-player all-pay auction. Player $i$ has a positive payoff if, and only if, there exists an $\bar{s}_i$ such that
    \begin{align*}
         \int_0^{\bar{s}_i}&\left( \frac{c'_{i}(x)}{v_{i}(x;x)} + \int_0^{x} c'_{i}(y) \frac{r_{i}(x;y)}{v_{i}(y;y)} \, dy \right)\, dx \\
         &<
         \int_0^{\bar{s}_i} \left(\frac{c'_{-i}(x)}{v_{-i}(x;x)} + \int_0^{x} c'_{-i}(y) \frac{r_{-i}(x;y)}{v_{-i}(y;y)} \, dy \right)\, dx = 1,
    \end{align*}
    where $ r_{i}(x;y)$, $ r_{-i}(x;y)$ are defined as in Theorem \ref{thm:main}.
    
    Moreover, Player $i$'s positive expected payoffs are given by:
    \[
    \left[1-\int_0^{\bar{s}_i}\left( \frac{c'_{i}(x)}{v_{i}(x;x)} + \int_0^{x} c'_{i}(y) \frac{r_{i}(x;y)}{v_{i}(y;y)} \, dy \right)\, dx\right]v_i(0;0). 
    \]
    
\end{corollary}

Corollary \ref{cor:payoff-iff} fully characterizes the payoffs of any two-payer all-pay auction with spillovers in terms of the model's primitives. While it is very general, it is not easily verifiable, justifying the use of simpler sufficient conditions. 

In contests without spillovers, as pointed out in \citet{siegel2009all,siegel2010asymmetric}, it is easy to identify the player with a positive payoff when normalized costs (i.e. the cost-value ratio) are ranked. That is, if, for all $s>0$,
\begin{equation}
    \frac{c_i(s)}{v_i(s)} < \frac{c_{-i}(s)}{v_{-i}(s)} \label{eq:normalizedc-nospillovers}
\end{equation}
holds in an auction with no spillovers, then player $i$ has a positive payoff.

This is consistent with Corollary \ref{cor:payoff-iff}. We can combine Corollaries \ref{cor:payoff-iff} and \ref{prop:bigG} to obtain the equivalent condition:

\[
    \frac{c_{i}(\bar{s})}{v_{i}(\bar{s}; \bar{s})} + 
    \int_0^{\bar{s}} \frac{\partial v_{i}(\bar{s}; y)}{\partial y} \frac{\tilde{G}_{-i}(y)}{v_{i}(\bar{s}; \bar{s})} dy
    <
    \frac{c_{-i}(\bar{s})}{v_{-i}(\bar{s}; \bar{s})} +
    \int_0^{\bar{s}} \frac{\partial v_{-i}(\bar{s}; y)}{\partial y} \frac{\tilde{G}_{i}(y)}{v_{-i}(\bar{s}; \bar{s})} dy
\]

In the absence of spillovers, the integral terms are equal to zero and the condition is implied by \eqref{eq:normalizedc-nospillovers}. In the presence of spillovers, one must impose a condition on the integrals.

\begin{theorem}\label{thm:syncomp}
    Consider a two-player all-pay auction where $v_i(s;y)$ is continuously differentiable in $s$ and $y$ for all $i \in I$. Suppose that the following two conditions hold:
    \begin{align}
        \frac{c_i(s)}{v_i(s; s)} &<  \frac{c_{-i}(s)}{v_{-i}(s; s)} \label{eq:sync1} \\
        \frac{1}{v_i(s; s)} \left| \frac{\partial v_i(s; y)}{\partial y} \right| &\leq \frac{1}{v_{-i}(s; s)} \frac{\partial v_{-i}(s; y)}{\partial y} \label{eq:sync2}
    \end{align}
    for all $s \in (0, \bar{s}]$ and $y \in [0,s]$. Then, Player $i$ has a positive payoff.

\end{theorem}

Theorem \ref{thm:syncomp} gives an analogue of \eqref{eq:normalizedc-nospillovers} for some contests with spillovers. The proof is in the Appendix. Condition \eqref{eq:sync1} is the same as \eqref{eq:normalizedc-nospillovers}, while Condition \eqref{eq:sync2} additionally imposes two extra requirements: (1) spillovers increase the value of the prize for Player $-i$ and (2) player $i$ is less dependant on these spillovers than her opponent.

Unlike contests without spillovers, condition \ref{eq:normalizedc-nospillovers} is not sufficient for the conclusion of Theorem \ref{thm:syncomp} to hold. Example \ref{example:rankedcost}, for instance, contains a situation where a player with strictly higher costs receives positive expected payoffs even when both players have exactly the same value function $v$ for the prize.

Corollary \ref{cor:identicalprize-payoff} shows that the negative effect of spillovers is indeed a necessary condition for a reversal in contests with symmetric prize values.

\begin{corollary}\label{cor:identicalprize-payoff}
    Consider a two-player all-pay auction with spillovers. Suppose the players have the same value $v(s; y) \equiv v_1(s; y) = v_2(s; y)$, which is continuously differentiable in both arguments and $c_2(s) > c_1(s)$ for all $s$. Then Player 2 has a positive payoff only if
    \[
        \frac{\partial v(s; y)}{\partial y} < 0
    \]
    for some $s,y$.
\end{corollary}

Intuitively, in these cases, a marginal increase in effort reduces the prize's attractiveness to the opponent by eroding its value. Thus, a player with higher absolute costs may nevertheless have lower marginal costs at value ranges where the value erosion inflicted on the opponent is substantial enough to suppress her incentives to win.

\pgfplotstableread{
xi    p1    p2
0.0000000000	0.0900180004	0.0000000000
0.0404040404	0.0873661202	0.0000000000
0.0808080808	0.0847387828	0.0000000000
0.1212121212	0.0820670404	0.0000000000
0.1616161616	0.0793935144	0.0000000000
0.2020202020	0.0767245316	0.0000000000
0.2424242424	0.0740668917	0.0000000000
0.2828282828	0.0713861962	0.0000000000
0.3232323232	0.0687301502	0.0000000000
0.3636363636	0.0660635328	0.0000000000
0.4040404040	0.0633923452	0.0000000000
0.4444444444	0.0607229764	0.0000000000
0.4848484848	0.0580621948	0.0000000000
0.5252525253	0.0554171348	0.0000000000
0.5656565657	0.0527952834	0.0000000000
0.6060606061	0.0502044673	0.0000000000
0.6464646465	0.0476031624	0.0000000000
0.6868686869	0.0449970525	0.0000000000
0.7272727273	0.0424436071	0.0000000000
0.7676767677	0.0399518034	0.0000000000
0.8080808081	0.0374240928	0.0000000000
0.8484848485	0.0349729498	0.0000000000
0.8888888889	0.0324974971	0.0000000000
0.9292929293	0.0300581135	0.0000000000
0.9696969697	0.0276619480	0.0000000000
1.0101010101	0.0253162833	0.0000000000
1.0505050505	0.0229695165	0.0000000000
1.0909090909	0.0206862708	0.0000000000
1.1313131313	0.0184132565	0.0000000000
1.1717171717	0.0162171116	0.0000000000
1.2121212121	0.0139800823	0.0000000000
1.2525252525	0.0118316196	0.0000000000
1.2929292929	0.0097149012	0.0000000000
1.3333333333	0.0076349117	0.0000000000
1.3737373737	0.0055966412	0.0000000000
1.4141414141	0.0036050724	0.0000000000
1.4545454545	0.0016651660	0.0000000000
1.4891022705    0.0000000000    0.0000000000
1.4949494949	0.0000000000	0.0001954029
1.5353535354	0.0000000000	0.0012407667
1.5757575758	0.0000000000	0.0022477244
1.6161616162	0.0000000000	0.0032555419
1.6565656566	0.0000000000	0.0042650581
1.6969696970	0.0000000000	0.0051953428
1.7373737374	0.0000000000	0.0061684532
1.7777777778	0.0000000000	0.0070611568
1.8181818182	0.0000000000	0.0079560942
1.8585858586	0.0000000000	0.0088541279
1.8989898990	0.0000000000	0.0097128774
1.9393939394	0.0000000000	0.0105755101
1.9797979798	0.0000000000	0.0113545897
2.0202020202	0.0000000000	0.0121820378
2.0606060606	0.0000000000	0.0129699690
2.1010101010	0.0000000000	0.0137178956
2.1414141414	0.0000000000	0.0144712376
2.1818181818	0.0000000000	0.0151844911
2.2222222222	0.0000000000	0.0158571933
2.2626262626	0.0000000000	0.0165360450
2.3030303030	0.0000000000	0.0172218323
2.3434343434	0.0000000000	0.0178673584
2.3838383838	0.0000000000	0.0184721742
2.4242424242	0.0000000000	0.0190845907
2.4646464646	0.0000000000	0.0196562006
2.5050505051	0.0000000000	0.0202360851
2.5454545455	0.0000000000	0.0207750492
2.5858585859	0.0000000000	0.0213229195
2.6262626263	0.0000000000	0.0218803609
2.6666666667	0.0000000000	0.0223460455
2.7070707071	0.0000000000	0.0228724814
2.7474747475	0.0000000000	0.0233062814
2.7878787879	0.0000000000	0.0238020529
2.8282828283	0.0000000000	0.0242043071
2.8686868687	0.0000000000	0.0246696851
2.9090909091	0.0000000000	0.0250406621
2.9494949495	0.0000000000	0.0254758479
2.9898989899	0.0000000000	0.0258157456
3.0303030303	0.0000000000	0.0262208709
3.0707070707	0.0000000000	0.0265840880
3.1111111111	0.0000000000	0.0269049478
3.1515151515	0.0000000000	0.0272378711
3.1919191919	0.0000000000	0.0275832912
3.2323232323	0.0000000000	0.0278861823
3.2727272727	0.0000000000	0.0281460941
3.3131313131	0.0000000000	0.0284745657
3.3535353535	0.0000000000	0.0287039984
3.3939393939	0.0000000000	0.0290027177
3.4343434343	0.0000000000	0.0292582787
3.4747474747	0.0000000000	0.0294702262
3.5151515152	0.0000000000	0.0296954054
3.5555555556	0.0000000000	0.0299341168
3.5959595960	0.0000000000	0.0301288596
3.6363636364	0.0000000000	0.0303372029
3.6767676768	0.0000000000	0.0305594041
3.7171717172	0.0000000000	0.0307372215
3.7575757576	0.0000000000	0.0308702028
3.7979797980	0.0000000000	0.0310757612
3.8383838384	0.0000000000	0.0312364717
3.8787878788	0.0000000000	0.0313518829
3.9191919192	0.0000000000	0.0315406862
3.9595959596	0.0000000000	0.0316841329
4.0000000000	0.0000000000	0.0317817744
}{\table}

\begin{figure}
    \centering
    \begin{tikzpicture} 
    \begin{axis}[
        xtick={0,1,1.4891022705,2,3,4},
        xticklabels={0,1,$\lambda^\star$,2,3,4},
        ytick=\empty,
        xlabel=Size of spillovers ($\lambda$),
        ylabel=Payoff,
        yticklabels={,,},
        legend pos=north east,
        width=0.8\textwidth,
        height=8cm,
    ]
    \addplot[smooth, thick] table [x = {xi}, y = {p1}] {\table};
    \addplot[smooth, thick, dashdotted] table [x ={xi}, y = {p2}] {\table};
    \legend{Player 1, Player 2}
    \end{axis}
    \end{tikzpicture}
    \caption{Player 2 has strictly higher costs. However Player 2 receives a positive payoff when spillovers are sufficiently large ($\lambda > \lambda^\star \approx 1.489$). }
    \label{fig:rankedc-parm}
\end{figure}
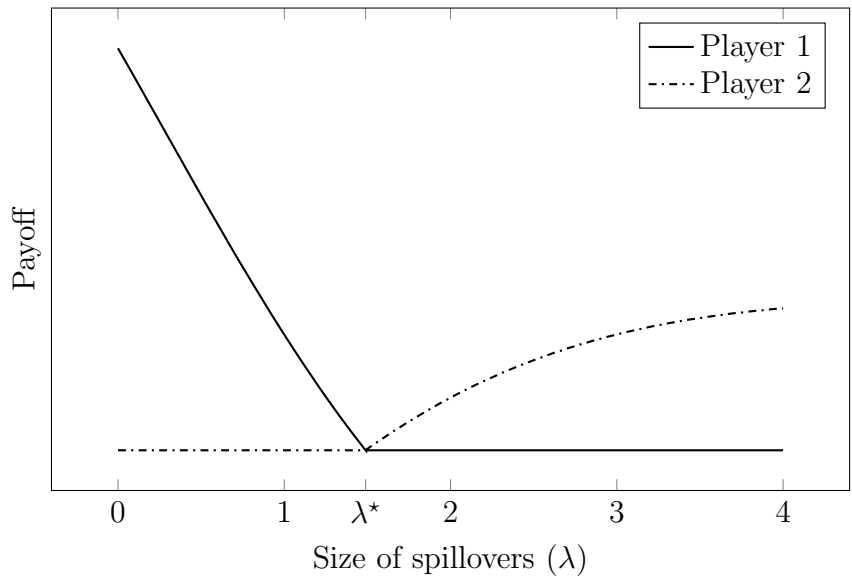

\begin{example}[Higher cost player has positive payoffs] \label{example:rankedcost} Consider a two-player contest with spillovers. Let $c_1(s) = s^2$,  $c_2(s) = s$, and $v(s;y):=v_1(s;y)=v_2(s;y)$ be given by:
\[
    v(s; y) = \frac{2}{5} + \frac{1}{1 + e^{\lambda(2y-1)}} 
\]
where $\lambda \geq 0$ is an exogenous parameter that determines the size of the spillovers. The support of the players strategies is contained in $[0,0.9]$. On this interval, Player 1 has a strict cost advantage.

When $\lambda = 0$, the prize is constant such that $v(s; y) = 0.9$. In this case, Player 1 receives a positive payoff. When we increase $\lambda$, this payoff decreases until we reach some $\lambda^\star\approx 1.489$ such that both players receive a payoff of zero. For all $\lambda > \lambda^\star$, 
Player 2 receives a positive payoff despite having strictly higher costs. The payoffs of both players are plotted in Figure \ref{fig:rankedc-parm}.

\end{example}

\begin{figure}
    \centering
\begin{tikzpicture} 
    \begin{axis}[
        domain=0:0.842019,
        samples=100,
        xtick={0,0.5,0.842019},
        xticklabels={0,0.5,$\bar{s}$},
        ytick=\empty,
        xlabel=$s_{-i}$,
        ylabel=$v_i(s_{-i})$,
        yticklabels={,,},
        legend pos=north west,
        width=0.5\textwidth,
        height=5cm
    ]
    \addplot [color=black, thick]    {0.4 + 1/(1 + e^(8*x - 4))};
    \end{axis}
\end{tikzpicture}
\begin{tikzpicture} 
    \begin{axis}[
        domain=0:0.842019,
        samples=100,
        xtick={0,0.5,0.842019},
        xticklabels={0,0.5,$\bar{s}$},
        ytick=\empty,
        restrict y to domain=0:1,
        xlabel=$s_i$,
        ylabel=$c_i(s_i)$,
        yticklabels={,,},
        legend pos=north west,
        width=0.5\textwidth,
        height=5cm
    ]
    \addplot [color=black, thick]    {x^2};
    \addplot [dashdotted, thick]  {x};
    \legend{$c_1$, $c_2$}
    \end{axis}
\end{tikzpicture}
\begin{tikzpicture} 
    \begin{axis}[
        domain=0:0.842019,
        samples=100,
        xtick={0,0.5,0.842019},
        xticklabels={0,0.5,$\bar{s}$},
        ytick=\empty,
        restrict y to domain=0:3.5,
        xlabel=$s_i$,
        ylabel=$g_i(s_i)$,
        yticklabels={,,},
        legend pos=north west,
        width=0.5\textwidth,
        height=5cm
    ]
    \addplot [color=black, thick] {1/(0.4 + 1/(1 + e^(8*x-4)))};
    \addplot [dashdotted, thick]  {2*x/(0.4 + 1/(1 + e^(8*x-4)))};
    \legend{$g_1$, $g_2$}
    \end{axis}
\end{tikzpicture}
\begin{tikzpicture} 
    \begin{axis}[
        domain=0:0.842019,
        samples=100,
        xtick={0,0.5,0.842019},
        xticklabels={0,0.5,$\bar{s}$},
        ytick={0,1},
        restrict y to domain=0:1,
        xlabel=$s_i$,
        ylabel=$G_i(s_i)$,
        yticklabels={,,},
        legend pos=north west,
        width=0.5\textwidth,
        height=5cm
    ]
    \addplot[smooth, thick] table [x = {s}, y = {G1}] {\table};
    \addplot[smooth, thick, dashdotted] table [x ={s}, y = {G2}] {\table};
    \legend{$G_1$, $G_2$}
    \end{axis}
\end{tikzpicture}
    \caption{Value (upper left), costs (upper right), strategy densities (lower left), and distributions (lower right) for Example \ref{example:rankedcost} with $\lambda = 4$. Note that Player 2 has a lower marginal cost for scores above $1/2$ and, because of spillovers, these scores devalue the prize for Player 1.}
    \label{fig:lambda4}
\end{figure}
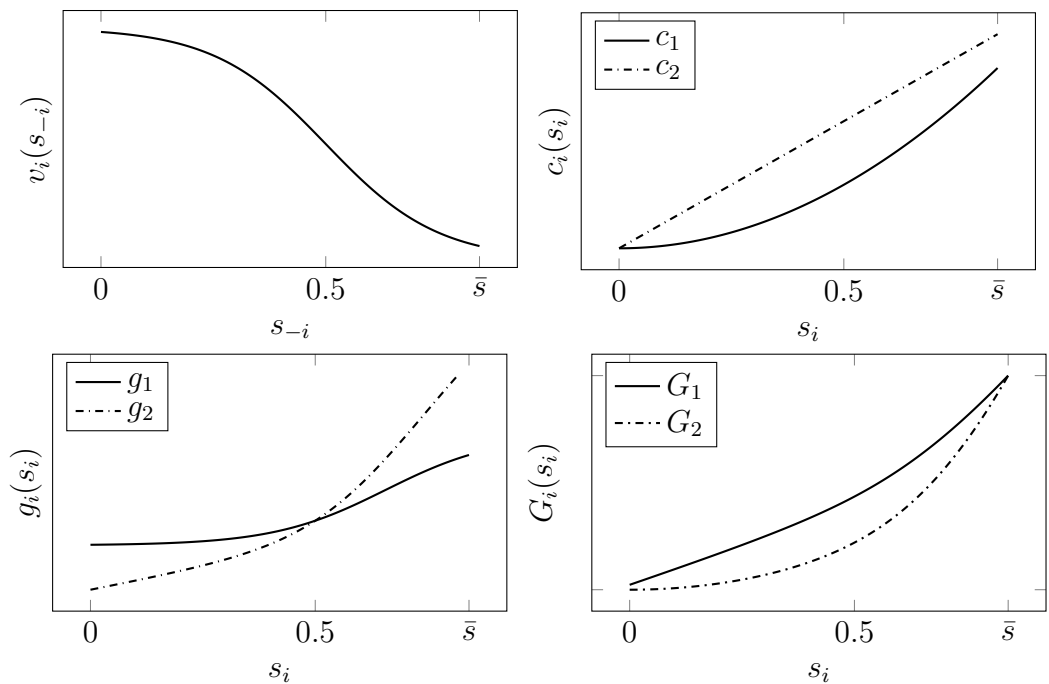

The reversal in Example \ref{example:rankedcost} occurs because marginal costs are not ranked. While Player 1 has lower costs in absolute terms, Player 2 has a lower marginal cost for all scores above $1/2$. This causes Player 2 to place comparatively more density on these bids. As can be seen in Figure \ref{fig:lambda4}, spillovers make the prize sharply less valuable when the opponent bids above $1/2$. So, these higher bids from player 2 damage player 1's valuation enough to reduce her participation.\footnote{The idea that underdogs may be harmed by policies that handicap stronger players appear in other settings in the literature. \citet{kirkegaard2013handicaps}, for example, shows that in contests with three players and incomplete information, handicapping the strongest contestant may indeed harm the weakest one. Interestingly, the fact that marginal costs are not ranked is also key in their construction: if the weakest player has a marginal cost advantage over low, but not high, scores, she might find it worthwhile to participate with low bids in the absence of handicaps.} 

Example \ref{example:rankedcost} highlights a potential problem when giving one side an advantage in a contest. In the presence of spillovers, decreasing a player's costs can reduce their welfare in equilibrium. The example also implies that it's possible to have a contest where one or more players would prefer to ex-ante increase their own costs.\footnote{Suppose both players are as in Example \ref{example:rankedcost} except $c_1(s) = c_2(s) = s^2$. Then, the game is symmetric. So, both players have a payoff of zero. If player $i$ increased her cost to $c_i(s) = s$, then she would receive a positive expected payoff, as in the example.}

\pgfplotstableread{
s    G1    G2
0.	0.0229702	0.
0.00842019	0.0290656	0.0000513318
0.0168404	0.0351666	0.000205454
0.0252606	0.0412736	0.000462573
0.0336808	0.047387	0.000822913
0.0421009	0.0535072	0.00128672
0.0505211	0.0596347	0.00185428
0.0589413	0.0657701	0.00252589
0.0673615	0.0719138	0.00330187
0.0757817	0.0780665	0.00418259
0.0842019	0.0842287	0.00516846
0.0926221	0.0904011	0.0062599
0.101042	0.0965843	0.00745739
0.109462	0.102779	0.00876146
0.117883	0.108987	0.0101727
0.126303	0.115207	0.0116917
0.134723	0.121442	0.0133191
0.143143	0.127692	0.0150558
0.151563	0.133958	0.0169024
0.159984	0.140241	0.0188599
0.168404	0.146542	0.0209293
0.176824	0.152863	0.0231116
0.185244	0.159205	0.0254078
0.193664	0.16557	0.0278194
0.202085	0.171958	0.0303475
0.210505	0.178371	0.0329937
0.218925	0.184812	0.0357594
0.227345	0.191281	0.0386465
0.235765	0.197781	0.0416568
0.244185	0.204314	0.0447921
0.252606	0.210881	0.0480548
0.261026	0.217486	0.0514471
0.269446	0.224129	0.0549715
0.277866	0.230815	0.0586307
0.286286	0.237545	0.0624277
0.294707	0.244323	0.0663655
0.303127	0.251151	0.0704475
0.311547	0.258032	0.0746774
0.319967	0.26497	0.0790589
0.328387	0.271968	0.0835963
0.336808	0.27903	0.0882939
0.345228	0.28616	0.0931566
0.353648	0.293361	0.0981894
0.362068	0.300638	0.103398
0.370488	0.307995	0.108787
0.378908	0.315436	0.114364
0.387329	0.322967	0.120135
0.395749	0.330593	0.126106
0.404169	0.338317	0.132285
0.412589	0.346147	0.13868
0.421009	0.354087	0.145299
0.42943	0.362143	0.152151
0.43785	0.370321	0.159243
0.44627	0.378626	0.166586
0.45469	0.387065	0.17419
0.46311	0.395645	0.182064
0.471531	0.404371	0.19022
0.479951	0.41325	0.198668
0.488371	0.422288	0.207421
0.496791	0.431492	0.216489
0.505211	0.440869	0.225885
0.513632	0.450425	0.235621
0.522052	0.460167	0.24571
0.530472	0.4701	0.256166
0.538892	0.480232	0.267001
0.547312	0.490569	0.278229
0.555732	0.501115	0.289863
0.564153	0.511878	0.301916
0.572573	0.522862	0.314402
0.580993	0.534073	0.327335
0.589413	0.545515	0.340727
0.597833	0.557193	0.354592
0.606254	0.569111	0.368943
0.614674	0.581272	0.38379
0.623094	0.593679	0.399148
0.631514	0.606335	0.415026
0.639934	0.619241	0.431436
0.648355	0.632399	0.448388
0.656775	0.645811	0.465892
0.665195	0.659475	0.483956
0.673615	0.673392	0.502589
0.682035	0.687562	0.521798
0.690455	0.701981	0.54159
0.698876	0.71665	0.561969
0.707296	0.731564	0.582942
0.715716	0.746722	0.604512
0.724136	0.762119	0.626682
0.732556	0.777751	0.649454
0.740977	0.793615	0.67283
0.749397	0.809705	0.696811
0.757817	0.826017	0.721396
0.766237	0.842544	0.746584
0.774657	0.859282	0.772375
0.783078	0.876223	0.798766
0.791498	0.893363	0.825754
0.799918	0.910694	0.853336
0.808338	0.928211	0.881507
0.816758	0.945906	0.910264
0.825178	0.963774	0.939602
0.833599	0.981807	0.969516
0.842019	1.	1.
}{\table}

Whenever marginal costs are ranked, the following proposition highlights necessary conditions for the high-marginal cost player to achieve a positive payoff. That is, the following conditions are necessary for a ``reversal'', where the higher cost player nonetheless obtains a positive payoff.

\begin{prop}[Ranked marginal costs]\label{prop:rankedmc}
    Consider a two-player all-pay auction. Suppose the players value for the prize is given by the function $v(s; y) := v_1(s; y) = v_2(s; y)$, and that $c'_2(s) > c'_1(s)$ for all $s\in S_i\cap S_{-i}$. Then Player 2 has a positive payoff only if all of the following apply
    \begin{enumerate}[label=(\roman*)]
        \item Costs are not scaled: there does not exist a $0<\lambda<1$ such that $c_1(s) = \lambda c_2(s)$ for all $s$.
        \item There exist some $t,z\in S_i\cap S_{-i}$ such that 
        \[
            \frac{\partial v(t; z)}{\partial z} < 0.
        \]
        \item There exist some $t,z \in S_i\cap S_{-i}$ such that $v'(t;z) > c'_2(t) - c'_1(t) > 0$.
        \item There exists some $s\in S_i\cap S_{-i}$ such that
        \[
            \max_{y \leq s} v'(s;y) - \min_{y \leq s} v'(s;y) > c'_2(s) - c'_1(s) > 0.
        \]
        \item There exist $t,z \in S_i\cap S_{-i}$ such that
        \[
           \frac{\partial v(t;z)}{\partial t \partial z} < 0,
        \]
        i.e. the common value function is not weakly supermodular.
    \end{enumerate}
\end{prop}

Proposition \ref{prop:rankedmc}, which is proven in the Appendix, is useful for contests where the opponent's actions are detrimental to the prize's value. It also pins down the circumstances under which the higher cost players can win: either the marginal costs are not ranked (as in Example \ref{example:rankedcost}) or all of the conditions 1-5 of Proposition \ref{prop:rankedmc} hold.

\section{Closed forms}\label{sec:closed-forms}

In some cases, it is possible to express the equilibrium strategies in closed form instead of as a series. We consider classes of prize value functions where this is possible. Section \ref{sec:other-applications} contains applications of each of the propositions below. 

\begin{prop}[Linearly separable spillovers] \label{prop:addsep}
Consider a two-player contest where $v'_{-i}(s_{-i} ; y)$ does not depend on $y$. That is, for each $i\in I$, $v_{-i}(s_{-i} ; y) = v^{-i}_{-i}(s_{-i}) + v^i_{-i}(y)$. Then,
\begin{equation}\label{eq:linearspill}
     \tilde{G}_i(s) =  
    \frac{1}{f(s)} \int_0^s \frac{c_{-i}'(y)}{v_{-i}(y; y)} f(y) dy,
\end{equation}
where 
\[
    f(y) := \exp \left(-\int _0^y \frac{ (v_{-i}^{-i})'(u)}{v_{-i}(u; u)} du\right).
\]
\end{prop}

When spillovers are not linear, we might still be able to find closed form solutions to equilibrium strategies. We highlight two particular cases where the VIEs in Equations \eqref{eq:ieq} and \eqref{eq:bigG} can be solved using Laplace transforms.

\begin{definition}[Laplace Transform]\footnote{See \citet{churchill1972operational} for an exposition on Laplace transforms.}
    A function $f$ defined on $\mathbb{R}_+$ admits a Laplace transform $F:\mathbb{C}\to\mathbb{C}$ given by
    \[
        F(x) := \mathcal{L} \{ f(s) \} = \int_0^\infty f(s) e^{-s x} ds
    \]
    if and only if the above integral conditionally converges. That is, it does not need to converge absolutely.
\end{definition}

We require an extra technical assumption to ensure that the integral above converges. For simplicity, we will assume the relevant functions are of exponential order.

\begin{definition}[Exponential order]\label{cond:existencelaplace} 
A function $f$ is of exponential order if and only if there exist $s'$, $q,M\in [0,\infty)$ such that, for all $s\geq s'$,
\[
    \left| f(s) \right|\leq Me^{qs}.
\]
\end{definition}

\begin{prop}[Margin of victory spillovers]\label{thm:laplace}
Assume a two-player contest such that \begin{enumerate*}[label=(\roman*)]
    \item for some $i$, $v'_i(s;y)/v_i(s;s)=:\nu_i(s-y)$  depends only on the score differential $s-y$, and
    \item $\nu_i$ and $c'_{-i}(s)/v_{-i}(s;s)$ are of exponential order. Though, it is sufficient to assume that it admits a Laplace transform.
\end{enumerate*} Then, for all $s\in (0,\overline{s}]$,
\[
    \tilde{g}_{-i}(s)= \mathcal{L}^{-1}\left\{\frac{\mathcal{L}\left\{\frac{c'_{i}(s)}{v_{i}(s;s)}\right\}}{1+\mathcal{L}\left\{{\nu}_{i}(s)\right\}}\right\}
\]
and
\[
    \tilde{G}_{-i}(s)= \mathcal{L}^{-1}\left\{\frac{\mathcal{L}\left\{\frac{c'_{i}(s)}{v_{i}(s;s)}\right\}}{x+x\mathcal{L}\left\{{\nu}_{i}(s)\right\}}\right\},
\]
where $\mathcal{L}$ and $\mathcal{L}^{-1}$ denote the Laplace and inverse Laplace transforms, respectively.
\end{prop}

Proposition \ref{thm:laplace} can be used whenever the prize's value depends on the margin of victory, i.e. on the difference $(s-y)$ between the winning bid $s$ and the losing bid $y$. We use Proposition \ref{thm:laplace} to solve the war of investment in Section \ref{sec:war_of_inv}.

\begin{prop}[Multiplicative margin of victory spillovers]\label{thm:laplace2}
    Assume a two-player contest such \begin{enumerate*}[label=(\roman*)]
        \item for some $i$, $(v_i(s;s))^{-1} (\partial v_i(s;y)/\partial y)=:\psi_i(s - y)$ depends only on the score differential $s-y$, and
        \item $\partial v_i(s;y)/\partial y$ and $\psi_i$ and $c_{i}(s)/v_{i}(s;s)$  are of exponential order.
    \end{enumerate*} Then, for all $s\in (0,\overline{s}]$,
    \[
        \tilde{g}_{-i}(s)= \mathcal{L}^{-1}\left\{\frac{x\mathcal{L}\left\{ 
        \frac{c_{i}(s)}{v_{i}(s;s)}
        \right\}}{1-\mathcal{L}\left\{\psi_{i}(s)\right\}}\right\}
    \]
    and
    \[
        \tilde{G}_{-i}(s)= \mathcal{L}^{-1}\left\{\frac{\mathcal{L}\left\{
        \frac{c_{i}(s)}{v_{i}(s;s)}
        \right\}}{1-\mathcal{L}\left\{\psi_{i}(s)\right\}}\right\}
    \]
    where $\mathcal{L}$ and $\mathcal{L}^{-1}$ denote the Laplace and inverse Laplace transforms, respectively.
\end{prop}

Proposition \ref{thm:laplace2} can be used whenever the prize value is of the form $ v_i(s;y) = v_i^1(s) v_i^2(s - y)$. For an application where we solve an all-pay auction with winner's regret, see Section \ref{sec:winners_regret}. 

\section{Applications}\label{sec:other-applications}

\subsection{War of attrition with costly preparation}\label{sec:WoAcp}

The canonical war of attrition is a game between two players $i=1,2$. Each picks a score, which represents an exit time, in $[0,\infty)$ and the player $i$ to select the largest score $s_i$ wins an amount that is decreasing in the loser's choice $s_{-i}$ and constant in her own. A player's payoff function is thus given by:
\[
    u_i(s_i; s_{-i}) =
    \begin{cases}
        f_i(s_{-i}) &\text{ if } s_i > s_{-i} \\
        \ell_i(s_{i}) &\text{ if } s_i < s_{-i} \\
        \alpha_i f_i(s_{-i}) + (1 - \alpha_i)  \ell_i(s_{i}) &\text{ if } s_i = s_{-i}
    \end{cases}
\]
where $f_i, \ell_i$ are strictly decreasing, continuously differentiable functions such that $f_i(s) > \ell_i(s)$, $\lim_{s\to\infty} \ell_i(s) = -\infty$, $\ell_i(0) = 0$, and $\alpha_i = 1 - \alpha_{-i} \in (0,1)$.

The typical WoA admits multiple equilibria and therefore does not satisfy the assumptions in Section \ref{sec:assumptions}. In particular, it violates monotonicity (A\ref{cond:mon}) and interiority (A\ref{cond:int}) because the payoff of the winner is constant (and therefore non-decreasing) in the Player's own score.  

We propose a general perturbation that selects a unique equilibrium of the WoA, and show that such a perturbation is solvable under our framework.\footnote{The problem of equilibrium selection in WoAs has been widely studied in the literature \citep{georgiadis2022absence, myatt2005instant}. One way to select a unique equilibrium is to truncate the game, as in \citet{nalebuff1985exit}, so that at some point in finite time both players prefer to exit. A different way to select for an equilibrium, which we discuss in more detail, is to introduce a small probability that a player that never exits. See, for example, \citet{abreu2000bargaining,kambe2019n,Kornhauser1989ReputationAP}.}  

Suppose, the winner's outcome is decreasing in her own score -- even if this dependence is minimal:
\[
    u_i(s_i; s_{-i}) =
    \begin{cases}
        f_i(s_{-i}) - \varepsilon_i(s_i) &\text{ if } s_i > s_{-i} \\
        \ell_i(s_{i}) - \varepsilon_i(s_i) &\text{ if } s_i < s_{-i} \\
        \alpha_i f_i(s_{-i}) + (1 - \alpha_i) \ell_i(s_{i}) - \varepsilon_i (s_{i}) &\text{ if } s_i = s_{-i}
    \end{cases}
\]
for any strictly increasing continuously differentiable function $\varepsilon_i$ with $\varepsilon_i(0) = 0$ and $\lim_{s\to\infty}\varepsilon_i(s)>f_i(0)$ for all $i$.

We denominate this variant a \textit{WoA with costly preparation}, as there is some small preparation cost $\varepsilon(s)$ incurred to set score $s$ -- i.e. the maximum amount of time $s$ one wishes to participate for. For example, a company engaged in a price war might have to build up inventory in advance or secure a costly line of credit. 

A WoA with costly preparation fits the two-player all-pay auction with spillovers where
\[
    v_i( s_i; s_{-i} ) :=  f_i(s_{-i}) - \ell_{i}(s_i) \text{ and}
\]
\[
    c_i(s_i) := \varepsilon_i(s_i) - \ell_i(s_i),
\]
which satisfy assumptions A\ref{cond:smo}-\ref{cond:pos}. Therefore, this game has a unique equilibrium, and there exists some $\bar{s}$ such that no player bids above $\bar{s}$. Theorem \ref{thm:main} further allows us to characterize the equilibrium and Proposition \ref{prop:addsep} gives a closed form expression for the equilibrium strategies.

As the preparation costs become small (with $\varepsilon'_i(s) \to 0$ uniformly for all $s$), the unique equilibrium of a WoA with costly preparation approaches the mixed-strategy equilibrium of the classic WoA. This is proven in the Appendix.

The WoA with costly preparation generalizes other perturbations that have a unique equilibrium. For example, \citet{abreu2000bargaining} and \citet{kambe2019n} extend the WoA to let a rational player's opponent be of an uncompromising type with positive probability, where ``uncompromising'' describes someone who bids (or exits at) infinity. Let $z_i$ denote the (known) probability that player $i$ is of an uncompromising type. Against such an opponent, a rational or compromising player loses with certainty. This is a special case of the WoA with costly preparation where $\varepsilon_i(s) := - (z_{-i}/(1-z_{-i})) \ell_i(s)$.

This relationship sheds light on the uniqueness of equilibrium found in the WoA with an uncompromising type. Indeed, by adding the possibility of a never-yielding opponent, we effectively introduce an unavoidable cost that depends on the player's own score. As was shown in the WoA with costly preparation, this characteristic is actually sufficient for a unique equilibrium.

\subsection{Offensive/defensive balance}\label{sec:off_def_balance}

    Military strategists generally agree that warfare is naturally asymmetric: the defending party can usually prevail with less expenditure of resources than the attacker \parencite{clausewitz1982war}. More generally, scholars have tried to identify which factors influence the so-called offensive/defensive balance -- that is, the many elements of military technology that generate either offensive or defensive advantages, and thus affect the probability of war \parencite{levy1984offensive}. Our model is able to capture both the defensive advantage and the role of the prize-depleting nature of war in the offensive/defensive balance debate.

    An attacker ($a$) invades a defender's ($d$) territory, which is worth $V$. Both combatants purchase costly scores in $[0,\infty)$, and the combatant with the higher score wins. A score of $s_i$ costs $c_i s_i$, where $c_i>0$ is a positive constant, for player $i\in I:=\{a,d\}$. Furthermore, $a$'s score inflicts $\delta_a s_a$ damage to the territory. Assuming the defender also inflicts a cost of $\delta_d s_d$ onto the attacker does not change the analysis. If the attacker wins, it internalizes all costs faced by the defender, as these costs effectively depleted the resources available from the territory. Consider the following payoff functions $u_a:[0,\infty)\to\mathbb{R}$ for the attacker:
    \begin{alignat*}{3}
        u_a(s_a, s_d) = p_a(s_a,s_d)(V - \delta_a s_a) &{- c_a s_a}, \\
    \intertext{and the following payoff function $u_d:[0,\infty)\to\mathbb{R}$ for the defender:}
        u_d(s_a, s_d) = (1-p_a(s_a,s_d))(V - \delta_a s_a) &{- c_d s_d},
    \end{alignat*}
    where $p_a(\cdot):[0,\infty)^2\to [0,1]$ denotes the probability that the attacker is victorious. Accordingly, we let $p_a(s_a,s_d) = 1$ whenever $s_a>s_d$, $p_a(s_a,s_d) = 0$ when $s_a<s_d$, and $p_a(s_a,s_d) = \lambda\in[0,1]$ whenever $s_a=s_d$. 

    When we transform this model into our framework, we get $c_i(s_i) := c_i s_i$ and
    \[
        v_a(s_a; s_d) = v_d(s_d; s_a) := V - \delta_a s_a.
    \]
    
    Assume it costs weakly more to attack than to defend (i.e., $c_a \geq c_d $). The attacker does not have any spillovers while the defender is harmed by her opponent.
    
    We are able to leverage the linearity of payoffs in this case to obtain a closed-form solution to the problem using Proposition \ref{prop:addsep}. The defender receives positive payoffs if, and only if,
    \[
        \bar{s}_d = \frac{V}{c_a + \delta_a} < \frac{V}{\delta_a} \left[ 1 - \exp{ \left( - \frac{\delta_a}{c_d} \right) } \right] = \bar{s}_a,
    \]
    which holds whenever $\delta_a > 0$ and $c_a \geq c_d$. 
    In this case,
    \[
        G_a(s) = 1 + \frac{c_d}{\delta_a} \log{\left[ \frac{c_a V}{(c_a + \delta_a) ( V - \delta_a s )} \right]}
        \quad  \text{ and } \quad
        G_d(s) = \frac{c_a s}{V - \delta_a s}.
    \]
    The probability $P(s_a>s_d|\delta_a,c_a,c_d)$ that the attacker succeeds,
    in equilibrium, is given by
    \[
        P(s_a>s_d|\delta_a,c_a,c_d) = \frac{c_d}{\delta^2_a} \left( \delta_a + c_a \log \left[ \frac{c_a}{c_a + \delta_a} \right] \right) < \frac{c_d}{2 c_a} \leq \frac{1}{2},
    \]
    where the supremum is reached as $\delta_a \to 0$. If the war damages the territory at least as much as it costs the attacker to inflict such damage, ($\delta_a \geq c_a$), a tighter bound is obtained:
    \[
        P(s_a>s_d|\delta_a,c_a,c_d) < 1 - \log{(2)} < \frac{1}{3}.
    \]
    Even if $c_a = c_d$, the defender is more than twice as likely to win than the attacker is. In our model, the stronger position of the defensive party comes as a byproduct of the inverse relationship between the attacker's strength and the erosion of the prize's value. This provides an alternate explanation on why it is typically easier to defend than to attack, something usually attributed to the high costs of maintaining long supply lines and of keeping seized territories \parencite{glaser1998what}. The defender's stronger position also suggests that any positive participation cost in a war contest imposed on the aggressor would be effective in discouraging aggression.
    \footnote{In the more general nonlinear model, where the value of the territory after invasion is given by $v_\delta(s_a)$ and the cost of choosing score $s_i$ to player $i$ is given by a continuously differentiable function $c_i:[0,\infty)\to\mathbb{R}_+$ satisfying the required assumptions A\ref{cond:smo} to A\ref{cond:pos}, $c_d(s) \leq c_a(s)$ is sufficient to ensure that $\tilde{G}_a(s) < \tilde{G}_d(s)$ for all $s>0$. This guarantees the defender's payoff remains positive, with $G_d(s) = c_a (s)/v_{\delta}(s)$.}

\subsection{War of investment} \label{sec:war_of_inv}
    Investment has long been considered as a method of committing to entry deterrence \parencite{dixit1980role}, while the war of attrition is a popular model of exit \parencite{fudenberg1986theory}. Our model can combine the two attributes into a single model of competition in continuous time, where players invest to stay in the game, but are able to recoup part of that investment if their opponent invests less. Wars of investment can also be used to model Cold-War style defense spending and competition between technology companies and R\&D races.
    
    Assume two competitors, $1$ and $2$, invest in capital $s_i$ at cost $c_i(s_i)$. The capital is necessary to engage in competition and depreciates at a constant rate. Competition results in zero profits. However, the winner is able to extract monopoly profits and benefits from the remaining capital according to an increasing function $v_i(s_i - s_{-i})$. More concretely, assume payoffs are
    \[
        u(s_i; s_{-i}) =
        \begin{cases}
            v_i(s_i - s_{-i}) - c_i(s_i) &\text{ if } s_i > s_{-i} \\
            - c_i(s_i) &\text{ if } s_i < s_{-i} \\
            \alpha_i v_i(0) - c_i(s_i) &\text{ if } s_i = s_{-i}
        \end{cases}
    \]
    for any $\alpha_i \in [0,1)$.

    If assumptions A1-4 are met, there is a unique equilibrium of capital investments in mixed strategies on finite support. Moreover, the equilibrium admits a closed-form solution by Proposition \ref{thm:laplace}.
    
\begin{example}

    Let $v_i(s;y):=e^{\rho_i(s-y)} \omega_i $ and $c_i(s) = e^{\rho_i s}-1$, where $\omega_i,\rho_i \in (0, 1)$ for each $i\in I:=\{1,2\}$. Then, 
    \[
        \tilde{g}_{i}(s)=
        \frac{\rho_{-i}}{\omega_{-i}}
    \]
    so the equilibrium strategies, excluding the possible mass point at zero, will be uniform with $\tilde{G}_{i}(s)= \left( \rho_{-i}/\omega_{-i}\right) s$. 
    
    The pair of ratios $\omega_i/\rho_i$ is therefore a sufficient statistic for the equilibrium of this game. Assume, without loss of generality that this ratio is weakly larger for Player 1. Then, the maximum duration of the game is Player 2's ratio $\bar{s} = \omega_2/\rho_2$.
    
    The equilibrium is fully characterized by the overall \textit{strength} of the players $\bar{s}$ and the \textit{competitive balance} $\delta := (\omega_2/\rho_2)/(\omega_1/\rho_1) \in (0,1]$.
    
    Because the strategies are uniform, Player 1's average commitment duration is half of the strength. Player 2 on the other hand has a mass point of size
    \[
        G_2(0) = 1 - \delta
    \]
    which decreases as the competition becomes more balanced.
    
    Overall, the conflict is expected to last for
    \[
        \mathbb{E}[ \min(s_1,s_2) ] = \int_0^{\bar{s}} (1 - G_1(y))(1-G_2(y)) dy = \frac{\delta \bar{s}}{3}
    \]
    total periods. The relationship between overall power and war duration is one to one. The duration is also increasing in the competitive balance. So, a large strength differential implies the conflict will typically be short-lived, whereas close contests can have delayed resolutions.
    
\end{example}

\subsection{All-pay auction with winner's regret} \label{sec:winners_regret}
    
    Winner's regret is the remorse that the winner has from spending more than is necessary to win a contest or auction. This phenomenon has mostly been studied in the context of winner-pay first-price, auctions \parencite{engelbrecht1989effect, filiz2007auctions}. We instead apply our framework to model winner's regret in an all-pay auction.
    
    Let each Player $i\in I:=\{1,2\}$ choose a score in $[0,\infty)$. Suppose $i$ values the prize at $\mu_i(s_i)[ 1 - h_i(s_i - s_{-i})]$, where $\mu_i(s_i)$ is the player's objective value of the prize and $h_i(s_i - s_{-i})$ is the share of the winnings that is unappreciated due to regret. Each player pays the cost $c_i(s_i)$ whether they win or lose. So payoffs are
    \[
        u(s_i; s_{-i}) =
        \begin{cases}
            \mu_i(s_i)[ 1 - h_i(s_i - s_{-i}) ] - c_i(s_i) &\text{ if } s_i > s_{-i} \\
            - c_i(s_i) &\text{ if } s_i < s_{-i}, \\
            \alpha_i \mu_i(s_i) - c_i(s_i) &\text{ if } s_i = s_{-i},
        \end{cases}
    \]
    for any $\alpha_i \in [0,1)$. We assume all functions are continuously differentiable with $c'(s) > 0$ and $h_i'(s) \geq 0$. Moreover, $\mu(0) > h(0) = c(0) = 0$ and $c'(s) > \mu'(s)$ for each $s$, so that lower bids are preferable even with no regret. Intuitively, the regret function, $h$, should not exceed one. Though, this is not a technical requirement. We can solve for the equilibrium with two Players using Proposition \ref{thm:laplace2} and we can extend this equilibrium the game with any number of identical players or prizes using Theorem \ref{thm:symmetric-nplayer}.

\begin{example}
    Let $\mu_i(s) := \omega_i \in \left( 0, \frac{1}{2} \right)$, $h_i(s) = s^2/2$ and $c_i(s) := s - s^2/2$ for $s \in \left[ 0, 1 \right]$. Then,
    \[
        \tilde{g}_{i}(s) =
        \frac{e^{-s}}{\omega_{-i}}.
    \]
    
    Without loss of generality, let $\omega_1 \geq \omega_2$, implying Player 1 receives a non-negative payoff. Player 1 will thus play a truncated exponential distribution with parameter $1$ and support $[0,-\log(1 - \omega_2)]$. Her expected score will be: 
    \[
     \mathbb{E}[s_1|\omega_1,\omega_2] = 1+\left(\frac{1-\omega_2}{\omega_2}\right)\log(1-\omega_2).
    \]
    which depends negatively on her opponent's payoff scaling factor $\omega_2$. This is lower than in the same game without regret.
     
    The player with zero expected payoffs will place a mass point at zero of size:
    \[
        G_2(0) = 
        1 - \frac{\omega_2}{\omega_1}
    \]
    which is exactly the same size as if there were no regret. Player 2 will have expected score
    \[
        \mathbb{E}[s_2|\omega_1,\omega_2] = \frac{\omega_2}{\omega_1} \left[ 1+\left(\frac{1-\omega_2}{\omega_2}\right)\log(1-\omega_2) \right]
    \]
    which is also less than in the same game without regret. The expected sum of the two scores score is:
    \[
        \mathbb{E}[s_1+s_2|\omega_1,\omega_2] = \left( 1 + \frac{\omega_2}{\omega_1} \right) \left[ 1+\left(\frac{1-\omega_2}{\omega_2}\right)\log(1-\omega_2) \right]
    \]
    which is decreasing in $\omega_1$ and increasing in $\omega_2$. In contests such as a labor tournaments, a large productivity differential between participants in the form of a high $\omega_1$ and low $\omega_2$ depresses aggregate effort. This is true in a contest with no spillovers, but the partial derivative of $\omega_1$ is larger in absolute value when there is regret. That is, the effect is exacerbated by the fact that the stronger player is penalized for winning by a large margin.
\end{example}

\section{More players}\label{sec:more_players}

In contests with spillovers and more than two players, many of the results considered here are violated. Existence still holds  (see \cite{olszewski2022equilibrium}), but uniqueness does not. Moreover, expected payoffs will now depend on which equilibrium is played.\footnote{This is also true of contests with no spillovers if monotonicity does not hold \citep[Example 2]{siegel2009all}.}

When the normalized costs are ranked, Theorem 2 in \citet{siegel2010asymmetric} and Theorem 2 in \citet{siegel2009all} show that only two players ever participate in the equilibrium of a contest for a single prize. This effectively collapses the problem into a two-player contest.

A version of this condition holds in our setting. We still require normalized costs to be ranked in some sense, but in a way that takes the spillovers into account.

\begin{theorem}\label{thm:non-participation} Assume $i,j$, $i\neq j$, are two of the $n>2$ players in a contest satisfying assumptions A\ref{cond:smo} to A\ref{cond:pos}. 
Suppose that Player $i$ has a positive payoff in the two-player contest where $i$ and $j$ are the participants, and that the following ``ranked costs'' condition holds for all $k\not\in\{i,j\}$, $s \in \tilde{S}_k$ ,$s_i\in\tilde{S}_i$ and $s_j\in \tilde{S}_j$
\begin{equation}\label{eq:rc3}
    \frac{c_k(s)}{v_k\left(s; \mathbf{s}_{\{i,j\}}\right)} \geq \frac{c_j(s)}{v_j\left(s; \mathbf{s}_{\{i\}}\right)},
\end{equation}
where $\mathbf{s}_H$ is a vector of opponent scores that is zero for all players not in set $H$. Then, there exists an equilibrium where only Players $i$ and $j$ participate.

\end{theorem}

To understand condition \eqref{eq:rc3}, consider the candidate equilibrium where Players $i$ and $j$ compete using their two-player strategies and Player $k$ does not participate. By not participating, Player $k$ earns a payoff of zero -- the same payoff as Player $j$. Condition \eqref{eq:rc3} says that if she enters, Player $k$'s normalized cost will be higher at every point than Player $j$'s already is. Therefore, her payoff from participating is at most zero (Player $j$'s payoff). So, there is no profitable deviation for any player.

Note that it is possible for $k \succ j$ and $j \succ k$ in the sense of \eqref{eq:rc3} when spillovers decrease the value of the prize. In this case, there are multiple equilibria where different pairs of players participate.

In the absence of spillovers, multiple equilibria also arise with three or more players. However, if payoffs are asymmetric, there can be at most one equilibrium where the support of each player's strategy is a union of intervals. Additionally, the payoffs of each player are consistent across all equilibria. Neither of these properties hold in contests with spillovers. Payoffs generally vary across equilibria in which different Players participate.

\subsection{Symmetric equilibria}

The same method used to find the equilibrium of two-player auctions with spillovers can be applied more generally to find symmetric equilibria of all-pay auctions with $n>2$ identical players and $m<n$ prizes, where the value of the prize for any given participant depends on their own score and on the score of the first runner-up (the player with the $m+1$-th highest bid). More specifically, each prize has value $v(s;y)$ where $s$ is the player's own score and $y$ is the score of the first runner-up. When there is only one prize, this amounts to saying that its value depends only on the two highest bids. Spillovers depend only on the score of the runner up in many games such as the all-pay auction with winner's regret and any game with a structure that resembles a war of attrition. For example, bargaining games and free riding games frequently have this structure where the last holdout to comply delays the prize for the winners. In the case where there is one prize, spillovers that depend on the first runner up capture the margin of victory which is relevant in many applications including elections and R\&D races.

Formally, we define a symmetric auction with runner-up spillovers as a family $\{I,P,\{\tilde{S}_i\}_{i\in I},\{u_i\}_{i\in I}\}$ where
\begin{enumerate}
    \item $I:=\{1,2,\dots,n\}$ is the index set of players, with $n\geq 2$.
    \item $P:=\{1,2,\dots,m\}$ is the index set of prizes, with $m<n$.
    \item For each $i\in I$, $\tilde{S}_i:=[0,\infty)$ is Player $i$'s action space. We let $\mathbf{s}_{-i}$ denote an arbitrary element of $\tilde{S}_{-i}:=\prod_{j\neq i}\tilde{S}_j$. We further let $s_{(j)}$ denote the $j$-th highest score.
    \item For each $i\in I$, $u_i:\tilde{S}\to \mathbb{R}$, where $\tilde{S}=\prod_{i\in I}\tilde{S}_i$.
    
    For each $\mathbf{s}:=(s_i;\mathbf{s}_{-i})\in \tilde{S}$, we further define:
    \[
    u_i(\mathbf{s}):= p_i(\mathbf{s})v(s_i,s_{(m+1)})-c(s_i)
    \]
    where (i) $p_i(\mathbf{s})$ denotes the probability that $i$ wins a prize given the score profile $\mathbf{s}$, with $\sum_{i\in I}p_i(\mathbf{s})=m$ and
\begin{align*}
     p_i(\mathbf{s}) = \left\{\begin{array}{ll}
       1 & \text{ if } s_{i} \geq s_{(m)}>  s_{(m+1)}, \\
        \frac{1}{|\{k\in I:s_k=s_{(m)}\}|} & \text{ if } s_{i} =  s_{(m)}=s_{(m+1)}, \\
        0 & \text{ if } s_{i} \leq s_{(m+1)} < s_{(m)};
    \end{array}\right.
\end{align*}
(ii)  $v:[0,\infty)^2\to\mathbb{R}_{+}$ maps each pair of scores $(s_i,s_{(m+1)})$  to Player $i$'s value $v( s_i;  s_{(m+1)} )$ from winning the prize, and (iii) $c:[0,\infty)\to\mathbb{R}_+$ outputs Player $i$'s private cost $c(s_i)$ given her submitted score $s_i$.
\end{enumerate}

In contrast with the two-player case introduced in Section \ref{sec:model}, here we assume all players are symmetric in the sense that they have identical value ($v$) and cost ($c$) functions. Moreover, all prizes are equally valuable to each player $i$ conditional on $(s_i,s_{(m+1)})$.

 In this context, we are able to use the two-player, one prize equilibrium characterized in Theorem \ref{thm:main} to construct the symmetric equilibria of a symmetric $n$-player, $m$-prize all-pay auction with spillovers
 
\begin{theorem}[Equilibrium of a symmetric $n$-player, $m$-prize all-pay auction with runner-up spillovers]\label{thm:symmetric-nplayer}
Consider a symmetric $n$-player, $m$-prize all-pay auction with runner-up spillovers. Assume $v,c$ satisfy assumptions A\ref{cond:smo} to A\ref{cond:pos}. Let $\hat{G}$ be defined as in Corollary \ref{prop:bigG}. That is, let $\hat{G}$ be the equilibrium cumulative distribution function of a two-player all-pay auction with spillovers:
\[
\hat{G}(s) = \frac{c(s)}{v(s,s)}+\int_{0}^s\frac{c(y)}{v(y,y)}\frac{R(s,y)}{v(s,s)}\,dy,
\]
with
\[
    R(s,y) = K^0(s,y)+K^1(s,y)+K^2(s,y)+\dots
\]
for $K^0(s,y)=\partial v(s,y)/\partial y$ and $K^t(s,y)$, $t=1,2,\dots$, defined recursively by $K^t(s,y):=\int_y^s (\partial v(s,z)/\partial z)(K^t(z,y)/v(z,z))\,dz$.

Then, the symmetric equilibrium of the $n$-player, $m$-prize all-pay auction with runner-up spillovers is given by the unique $G$ that solves:
\begin{equation}\label{eq:orderstat}
    \hat{G}(s) = \sum_{j=n-m}^{n-1} \binom{n-1}{j} [ G(s) ]^{j} [ 1 - G(s) ]^{n-j-1}.
\end{equation}
\end{theorem}

To see why Theorem \ref{thm:symmetric-nplayer} holds, consider the expected payoff of a Player $i$ who bids $s$:
\[
    \int_0^s v(s;y) \,d\hat{G}(y) - c(s),
\]
where $\hat{G}$ is the probability measure of the $m$-th largest score out of the $n-1$ players in $I\setminus\{i\}$. In a symmetric equilibrium, $\hat{G}$ is the $n-m$-th order statistic of a sample of $n-1$ draws from the equilibrium distribution $G$, giving us \eqref{eq:orderstat}. The right-hand side of \eqref{eq:orderstat} is increasing in $G(s)$, and thus may be inverted to obtain $G$ given $\hat{G}$.
 
At the same time, $\hat{G}$ is the equilibrium of a symmetric two-player auction, since each Player's indifference condition is identical to \eqref{eq:ieq}. This allows us to use Theorem \ref{thm:main} to find $\hat{G}$.
 
Theorem \ref{thm:symmetric-nplayer} shows that the equilibrium in the two-player case is also the symmetric equilibrium of the game with any number of players and prizes, subject to a particular monotone transformation. Intuitively, increasing the number of players (or decreasing the number of prizes) reduces the scores of each player.

Asymmetric equilibria are more difficult to characterize when either spillovers or incomplete information are present. In a world without spillovers but complete information, the equilibrium of an $n$-player, $m$-prize contest is unique under mild conditions, and \citet{siegel2010asymmetric} was able to provide an algorithm for its construction. Under incomplete-information, asymmetric equilibria with more than two players are notoriously difficult to analyze. A general characterization is still an open question, as far as we are aware.\footnote{See \citet{kirkegaard2013handicaps}, \citet{parreiras2010contests} for analyses on particular equilibria of $N\geq 3$-players all-pay, incomplete-information auctions.}

\section{Related literature and conclusions}\label{sec:conclusion}

Throughout this paper, we characterized and established uniqueness for the equilibrium of two-player contests using techniques from the theory of integral equations. We then extended these results to symmetric contests with more players and prizes to characterize the symmetric equilibrium. This allowed us to derive insights on equilibrium payoffs, winners and losers, and on the importance of spillovers for applications. This model does not require or imply that the results of a contest are known in advance. In fact, players are always uncertain of their own victory. However, this uncertainty stems from not knowing the resources that your opponent dedicated to the contest. The fact that ranked normalized costs are not enough to establish dominance demonstrates how spillovers can favor high-cost, low-value players that nevertheless have a marginal cost advantage over their opponent when bids are high. In particular, the results in this paper suggest several potential consequences of legal structures, conflicts and competition. 

This paper is most closely related to two others. \citet{baye2012contests}, also considers spillovers in two-player contests, but focuses on symmetric equilibria and linear symmetric costs and valuations. We show that there are no asymmetric equilibria in this two player case and extend the analysis to include asymmetric players and general functional forms for the prize values. This allows us to establish equilibrium uniqueness, express novel results about payoffs, and characterize the equilibrium in different applications (Sections \ref{sec:WoAcp} and \ref{sec:other-applications}).

The second paper that approaches a similar question to our own is \citet{xiao2018all}. The author, however, focuses on constant prize values and separable spillovers in the cost functions, which are independent of winning or losing. This independence significantly restricts the equilibrium effects of the spillovers. Linearly separable spillovers on the cost have no effect on the equilibrium, while multiplicatively separable spillovers scale the cost of bids by an endogenous constant. This is not true when spillovers are in the prize value.

\citet{fu2013competitive} also analyze two-player all-pay auctions with linear spillovers in the costs. They assume that each contestant is a firm with a minority stake in their opponent's profits. As such, even when a firm loses the auction, they still get to keep a share of the prize. On the other hand, regardless of winning or losing, they must also share in on the cost of effort incurred by their opponent -- hence, the existence of cost spillovers. Because these spillovers are linear and do not affect the prize of the winner, they have no effect on the equilibrium distributions -- as was also noted in \citet{xiao2018all}. 

Our paper is also connected more broadly to the literature of spillovers in other auction and auction-like frameworks. \citet{hodler2012all}, for example, use a linear first-price auction with spillovers to model war. The authors refer to this as an all-pay contest, but only the winner actually pays because of the way funds are handled.

Notably, spillovers have been given comparatively more attention in the Tullock contest framework. In these contests, each participants' probability of winning is given by $p_i(s_i;s_{-i})=s_i^r/(s_i^r+ s_{-i}^r)$,  if $(s_1,s_2)\neq (0,0)$, and $p_i(s_i;s_{-i})=1/2$ if $(s_1,s_2) = (0,0)$. Here, $r\in(0,\infty)$ is a parameter that controls how much one's probability of winning responds to an increase in scores. The all-pay auction is a Tullock contest where $r=\infty$; when $r=1$ we have a Tullock lottery instead.

\citet{chowdhury2011generalized} study a generalized Tullock lottery in which payoffs linearly incorporate one's own effort and the effort of the rival. The paper studies symmetric payoff and cost structures and, as is usual in Tullock-type contests, both players are able to extract positive payoffs. The authors obtain asymmetric, pure-strategy equilibria -- even when players are identical \citep{chowdhury2011multiple}.
In contrast, our all-pay framework yields a unique symmetric mixed-strategy equilibrium when players are identical, and expected payoffs are zero.

\citet{damianov2018asymmetric} allows for players' efforts to produce either positive (productive) or negative (destructive) externalities in a two-player Tullock lottery. The author finds that spillovers can either accentuate or reduce the competitive balance between participants, when contrasted to a comparable fixed-prize contest. However, unlike our results, no reversal can ever occur: the ``favored'' player is always more likely to win and has a higher expected payoff.\footnote{There are many other examples of spillovers in Tullock-type contests, see e.g. \citet{chung1996rent}, \citet{hirai2013existence}.}

We identify several avenues for future work. The class of contests that include spillovers is very large and fits many applications. The fact that we are able to construct very different contests with the same equilibrium strategies (e.g. the all-pay auction with winner's regret in Section \ref{sec:winners_regret} has the same equilibrium as a war of attrition with costly preparation of Section \ref{sec:WoAcp}) suggest that it might be possible for a contest designer to induce behavior more cheaply through spillovers.

Other contest design problems where spillovers are available are also of interest. \ref{appendixopt} contains a brief exposition that shows that when a constrained designer that cares about aggregate effort can reward contestants with prizes that may include spillovers, no contestant will be allowed positive rents. This in particular would make computing equilibrium strategies straightforward. Under what circumstances introducing spillovers is desirable to a contest designer is however still an open question.

\printbibliography

\section{Appendix}

\renewcommand{\thesubsection}{Appendix \Roman{subsection}}

\subsection{Proofs}

\paragraph{Proof of Lemma \ref{prop:interval}}

\newcommand{\lemtemp}{\thelemma}
\setcounter{lemma}{-1} 
\begin{lemma}(Interval support)\label{prop:interval}
    In any Nash equilibrium, players choose strictly increasing, continuous mixed strategies $G_i$ with common support on some interval $[0, \bar{s}]$. At most one participant can have a mass point and it must be at zero.
    
    Moreover, the equilibrium is defined by the following indifference condition:
    \begin{equation}
        \tilde{g}_{i}(s) = \frac{c'_{-i}(s)}{v_{-i}(s; s)} - \int_0^s \frac{v'_{-i}(s; y)}{v_{-i}(s; s)} \tilde{g}_{i}(y) dy. \tag{\ref{eq:ieq}}
    \end{equation}
\end{lemma}
\setcounter{lemma}{\lemtemp} 
\begin{proof}
    
    The argument is standard in the all-pay auction literature, and is presented here for completeness. Our proof is in several steps.
    
    \begin{enumerate}
        \item By A\ref{cond:int}, each player $i\in I$ would select scores in $S_i=[0,T_i]$, for $T_i$ finite.
        
        \item By A\ref{cond:smo} and A\ref{cond:pos}, both players prefer to win than lose in a neighborhood of the opponent's score. This is used in points 3, 5, and 7.
        
        \item \textit{The minimum score in the support of both players' strategies is zero.} 
        
        Let $\underline{s}_1,\underline{s}_2$ denote the lower bounds of player $1$ and player $2$'s strategies' supports, respectively. Suppose $\underline{s}_i \geq \underline{s}_j$. Then, if $i$ places no atom in $\underline{s}_j$, $\underline{s}_j=0$ and $j$ will never want to play anything in $(0,\underline{s}_i)$ by A\ref{cond:mon}. If $i$ has no atom at $\underline{s}_i$, then $j$ won't want to play anything on the $(0,\underline{s}_i]$.
        
        Now suppose $\underline{s}_i > \underline{s}_j=0$ -- we will show that one of the two players at least has a profitable deviation. If $i$ has an atom at $\underline{s}_i$, she could bring the bottom of her support closer to zero, increase her payoffs, and not change her probability of winning. If $i$ does place an atom at $\underline{s}_i$ and $j$ does not, then she could move that point-mass at $\underline{s}_i$ closer to zero, spend less on bids and not change her probability of winning. Finally, if both $i$ and $j$ place an atom at $\underline{x}_i$, then (by A\ref{cond:smo}) either $i$ could do better by spreading that atom to a $\varepsilon$-neighborhood just above it (if $\underline{s}_i<1$), or $j$ would prefer to place that mass at 0 instead of $\underline{s}_i<1$ (if $\underline{s}_i=1$).
        
        Thus, it must be that $\underline{s}_i = \underline{s}_j=0$.
        
        \item \textit{Both players will have the same maximum score in their strategies' support ($\bar{s}$)}. Otherwise, the player with the highest upper bound to her support could reduce it and increase her payoff (by A\ref{cond:mon}) without impacting her probability of winning.
        
        \item \textit{There are no mass points on the half open interval $(0,\bar{s}]$.} If $i$ places a mass point at $s_i \in (0,\bar{s})$, then $j$ would find it worthwhile to transfer mass from a neighborhood below $s_i$ to one just above $s_i$. If $i$ places a mass point at $\bar{s}$, then $j$ would find it worthwhile to transfer mass from a neighborhood below $s_i$ to $0$. Either way, there would be an $\varepsilon$-neighborhood below $s_i$ in which $j$ would put no mass. But then it can't be an equilibrium strategy for $i$ to place an atom at $s_i$ in the first place.
        
        \item \textit{There are no gaps in the density.} Suppose that there is an interval $(s',s'') \subset [0,\overline{s}]$ where player $i$ places no probability mass. Pick this interval such that $s''$ is the ``largest'' point without density \footnote{Such a $s''$ exists since there are no mass-points in $i$'s distribution, implying it is continuous.} This is just to make sure there's some probability mass in the interval just above $s''$.

        We have that $j$ can't have any density on $(s',s'')$ either, or she would rather transfer it all to $x'$. But then, for $i$, anything too close to $s''$ from above is worse than picking $s'$.
        
        \item \textit{At most one player will place a mass point at zero.} The two players can't both have a mass point at $\underline{s}=0$: either player would rather move that mass infinitesimally above it.
    \end{enumerate}
    
    Therefore, there are mixed strategies on some interval $[0, \bar{s}]$ where the following indifference condition must hold for every point in the support of Player $i$:
    \begin{equation}
        \bar{u}_i(G_{-i}) := \int_{0}^{s}v_i(s;y) \, dG_{-i}(y) - c_i(s) \qquad \text{ for }s\in[0,\bar{s}]. \tag{\ref{eq:ind}}
    \end{equation}
    We can uniquely decompose $G_{-i}$ into a continuous measure $\tilde{G}_{-i}$ and a mass-point measure using the Lebesgue decomposition. Because there can be only one mass point at zero, we decompose to
    \[
        \bar{u}_i(G_{-i}) = \int_0^s v_{i}(s; y) \, d \tilde{G}_{-i}(y) +  v_{i}(0,0) G_{-i}(0) - c_{i} (s).
    \]
    Plugging in $s=0$ reveals $\bar{u}_{i} = v_{i}(0,0) G_{-i}(0)$. So, the condition simplifies to
    \[
        c_{i} (s) = \int_0^s v_{i}(s; y) \, d \tilde{G}_{-i}(y).
    \]
    Therefore, the right hand side is differentiable on the support of Player $i$. This implies that
    \begin{align*}
        c'_{i} (s) &= \lim_{\varepsilon \to 0} \varepsilon^{-1} \left[ \int_0^{s+\varepsilon} v_{i}(s + \varepsilon; y) \, d \tilde{G}_{-i}(y) - \int_0^s v_{i}(s; y) \, d \tilde{G}_{-i}(y) \right] \\
                    &= \int_0^s v'_{i}(s; y) \, d \tilde{G}_{-i}(y) + v_{i}(s ; s) \lim_{\varepsilon \to 0} \varepsilon^{-1} \left[ \int_s^{s+\varepsilon}  \, d \tilde{G}_{-i}(y) \right]
    \end{align*}
    holds. Then, $v_{i}(s ; s) \neq 0$ implies that the limit exists and is equal to
    \begin{align*}
         \lim_{\varepsilon \to 0} \varepsilon^{-1} \left[ \tilde{G}_{-i}(s + \varepsilon) - \tilde{G}_{-i}(s) \right] &= \lim_{\varepsilon \to 0} \varepsilon^{-1} \left[ \int_s^{s+\varepsilon}  \, d \tilde{G}_{-i}(y) \right] \\
         &= \frac{c'_{i} (s)}{v_{i}(s ; s)} - \int_0^s \frac{v'_{i}(s; y)}{v_{i}(s ; s)} \, d \tilde{G}_{-i}(y).
   \end{align*}
    Then, $\tilde{G}_{-i}$ is differentiable on Player $-i$'s support with a continuous derivative. 
    
    Consider the case where Player $i$ plays with full support on $[0, \bar{s}]$. Because continuous differentiability implies absolute continuity on a bounded interval, the above can be rewritten as
    \begin{equation}
        \tilde{g}_{-i}(s) = \frac{c'_{i}(s)}{v_{i}(s; s)} - \int_0^s \frac{v'_{i}(s; y)}{v_{i}(s; s)} \tilde{g}_{-i}(y) dy. \tag{\ref{eq:ieq}}
    \end{equation}
    
    Now, suppose that there are some points not on the support such that \eqref{eq:ind} does not hold with equality. If there exists a positive $\tilde{g}_{-i}$ that solves \eqref{eq:ieq}, then at any such point, $t$,
    \[
        \int_{0}^{t} v_i(t;y) \tilde{g}_{-i}(y) \, dy = c_i(t) \geq \int_{0}^{t} v_i(t;y) \, d\tilde{H}_{-i}(y).
    \]
    The Lebesgue decomposition ensures that $\tilde{H}_{-i}$ can be broken down to an absolutely continuous portion (which must agree with $\tilde{G}_{-i}$) and a continuous singular portion. Therefore, we can equivalently write
    \[
        0 \geq \int_{0}^{t} v_i(t;y) \, d\hat{H}_{-i}(y).
    \]
    where $\hat{H}_{-i}$ is the singular part of $\tilde{H}_{-i}$.
    
    We now show that $\hat{H}_{-i}(s) = 0$ for all $s \in [0, \bar{s}]$. Suppose, by way of contradiction that there exists an $x$ such that $\hat{H}_{-i}(x) > 0$. By continuity, $\hat{H}_{-i}(0)=0$.
    
    Because $\hat{H}_{-i}$ is continuous, there must exist a point $t$ such that $\hat{H}_{-i}(t) = 0$ and $\hat{H}_{-i}(t+\varepsilon) > 0$ for all $\epsilon > 0$. Then,
    \begin{align*}
        0 &\geq \int_t^{t+\varepsilon} v_{-i}(t + \varepsilon; y) \, d Q_i(y) \\
          &\geq \left[ \min_{y \in [t,t+\varepsilon]} v_{-i}(t; y) \right] \left[ \hat{H}_{-i}(t+\varepsilon) - \hat{H}_{-i}(t) \right] \\
          &= \left[ \min_{y \in [t,t+\varepsilon]} v_{-i}(t; y) \right] Q_i(t+\varepsilon).
    \end{align*}
    Therefore, for any $\varepsilon > 0$, 
    \[
        \min_{y \in [t,t+\varepsilon]} v_{-i}(t; y) \leq 0.
    \]
    However, $v_{-i}(t; y)$ is continuous and $v_{-i}(t; t) > 0$. This is a contradiction.
\end{proof}

\paragraph{Proof of Lemma \ref{prop:pdf}}

\begin{proof}
    The finite definite integral cannot diverge because the function is continuous. Also note that \eqref{eq:ieq} gives us $g_i(0) = c'_{-i}(0)/v_{-i}(0; 0) \geq 0$. This inequality is strict if $c'_{-i}(0) > 0$.
    
    If the inequality is not strict, we find a positive value in a neighborhood of zero. Because $c'_{-i}$ is positive everywhere but zero, we know that $g_i$ is strictly increasing near zero. So there exists some $\delta > 0$ such that $g_i(s) > 0$ for $s \in (0, \delta)$.
    
    We still need to confirm that $\tilde{g}_i(s) > 0$ on the relevant interval $ \{ s : \int_0^s | \tilde{g}_i(y) | dy \leq 1 \}$. Suppose, by way of contradiction, that it is not. Then, by continuity, there must be an initial point $s^\star > 0$ such that $\tilde{g}_i(s^\star)=0$, $ \int_0^{s^\star} \tilde{g}_i(y) dy \leq 1 $, and $\tilde{g}_i(s) \geq 0$ for all $s \leq s^\star$. However, this is impossible because
    \begin{align*}
        \tilde{g}_i(s^\star) &= \frac{1}{v_{-i}(s^\star; s^\star)} \left( c'_{-i}(s^\star) - \int_0^{s^\star} v'_{-i}(s^\star , y) |\tilde{g}_i(y)| dy \right) \\
        &\geq \frac{1}{v_{-i}(s^\star; s^\star)} \bigg[ c'_{-i}(s^\star) - \left| \max_{y\in[0,s^\star]} v'_{-i}(s^\star; y) \right| \underbrace{\left( \int_0^{s^\star} |\tilde{g}_i(y)| dy \right)}_{\leq 1} \bigg] \\
        &\geq \frac{1}{v_{-i}(s^\star; s^\star)} \underbrace{\left[c'_{-i}(s^\star) - \left| \max_y v'_{-i}(s^\star; y) \right|  \right]}_{>0 \text{ (A\ref{cond:mon})}} > 0.
   \end{align*}
   
   We must now show that it is not possible for $\int_0^\infty | \tilde{g}_i(y) | dy \leq 1$. We can do this in one step with Holder's inequality.
   \[
   c_{-i}(s) = \int_0^s v_{-i}(s; y) g_i (y) dy \leq \left(\int_0^s |g_i(y)| dy\right) \left( \max_{y \in [0,s]} v_{-i}(s; y) \right)
   \]
   so $\int_0^s |g_i(y)| dy \geq c_{-i}(s)/(\max_y v_{-i}(s; y))$ which is assumed to be greater than one as $s$ approaches infinity (A\ref{cond:int}). By continuity, there exists an $\bar{s}_i$ such that $\int_0^{\bar{s}_i} |g_i (y)| dy = 1$ (A\ref{cond:smo}).
\end{proof}

\paragraph{Proof of Corollary \ref{prop:bigG}}
\begin{proof}
    Equation \eqref{eq:bigG} is obtained by applying integration by parts to \eqref{eq:ind}. This defines a Volterra Integral Equation which has a unique solution by lemma \ref{prop:invert}. This solution coincides with the one in Theorem \ref{thm:main} because Equation \eqref{eq:ind} cannot have two solutions.
\end{proof}

\paragraph{Proof of Theorem \ref{thm:syncomp}}

\begin{proof}
    Consider equation \eqref{eq:bigG}. The main result of \cite{beesack1969comparison} allows us to compare the solutions of two VIEs. In our setting, this means that conditions \eqref{eq:sync1}, \eqref{eq:sync2} imply
    \[
        \tilde{G}_2(s) \leq \tilde{G}_1(s) + \frac{c_1(s)}{v_1(s; s)} - \frac{c_2(s)}{v_2(s; s)} < \tilde{G}_1(s).
    \]
    From this, it is clear that $\bar{s}_1 \leq \bar{s}_2$ which implies that player 2 has a mass point. The bound comes from
    \[
        u_1 = v_1(0; 0) (1 - \tilde{G}_2(\bar{s})) \geq v_1(0; 0) \left[ \frac{c_2(\bar{s})}{v_2(\bar{s}; \bar{s})} - \frac{c_1(\bar{s})}{v_1(\bar{s}; \bar{s})} \right].
    \]
\end{proof}

\paragraph{Proof of Corollary \ref{cor:identicalprize-payoff}}

\begin{proof}
We want to apply Theorem \ref{thm:syncomp}. Because the prizes are identical, \eqref{eq:sync1} is satisfied by the ranked cost assumption and \eqref{eq:sync2} holds with equality as long as the derivative is non-negative.
\end{proof}

\paragraph{Proof of Proposition \ref{prop:rankedmc}}

We first prove that it's not possible to have a reversal when costs are scaled (Condition 1).

\begin{proof}
    Suppose $v(s ; y) := v_1(s ; y) = v_2(s ; y)$ and $c_2(s) = \lambda c_1(s)$ where $\lambda > 1$. Because the two players have the same kernel, they must share the same resolvent ($R$). Then,
    \begin{align*}
        \tilde{G}_1 (\bar{s}) &= \frac{c_2(\bar{s})}{v(\bar{s}; \bar{s})} + \int_0^{\bar{s}} R(\bar{s}, y) \frac{c_2(y)}{v(y; y)} dy \\
        &= \lambda \left( \frac{c_1(\bar{s})}{v(\bar{s}; \bar{s})} + \int_0^{\bar{s}} R(\bar{s}, y) \frac{c_1(y)}{v(y; y)} dy \right) \\
        &= \lambda \tilde{G}_2(\bar{s}),
    \end{align*}
    implying $\tilde{G}_1 (\bar{s}) > \tilde{G}_2 (\bar{s})$ and thus Player 2 must have a mass point at zero.
\end{proof}

Condition 2 is Corollary \ref{cor:identicalprize-payoff}. We then show that a reversal is impossible when $v'(t;z) \leq c'_2(t) - c'_1(t)$ for all $t,z$ (Condition 3).

\begin{proof}
    We would like to show that $\tilde{g}_1(s) - \tilde{g}_2(s) > 0$ for all $s \in [0, \bar{s}]$. This would imply that $\tilde{G}_1 (\bar{s}) > \tilde{G}_2 (\bar{s})$, which implies that Player 2 must have the mass point at zero. By \eqref{eq:ieq}:
    \[
        \tilde{g}_{1}(s) - \tilde{g}_{2}(s) = \frac{c'_{2}(s) - c'_{1}(s)}{v(s; s)} - \int_0^s \frac{v'(s; y)}{v(s; s)} \left( \tilde{g}_{1}(s) - \tilde{g}_{2}(s) \right) dy.
    \]
    This is positive by Lemma \ref{prop:pdf}.
\end{proof}

Finally, we prove Conditions 4 and 5 by contradiction. The two proofs share the same initial setup.

\begin{proof}
    Suppose, by way of contradiction that Player 1 does not have a positive payoff. Then, there exists an $\bar{s}$ such that $\tilde{G}_2(\bar{s}) = 1$ and $\tilde{G}_1(\bar{s}) \leq 1$. Note that $\tilde{G}_1(0)-\tilde{G}_2(0) = 0$ and
    \[
        \tilde{g}_1(0)-\tilde{g}_2(0) = \frac{c'_2(0)-c'_1(0)}{v(0;0)} > 0.
    \]
    Therefore, there exists some $r \in (0, \bar{s}]$ that is the first point such that $\tilde{G}_1(r)-\tilde{G}_2(r) = 0$ and $\tilde{g}_1(r)-\tilde{g}_2(r) \leq 0$. 
    
    If we assume $c'_2(r) - c'_1(r) > \max_{y \leq r} v'(r;y) - \min_{y \leq r} v'(r;y)$, then
    \begin{align*}
        \tilde{g}_1(r)-\tilde{g}_2(r) &= \frac{c'_2(r)-c'_1(r)}{v(r;r)} - \int_0^{r} \frac{v'(r;y)}{v(r;r)} [\tilde{g}_1(y)-\tilde{g}_2(y)] dy \\
        &= \frac{c'_2(r)-c'_1(r)}{v(r;r)} - \int_0^{r} \frac{v'(r;y)}{v(r;r)} \tilde{g}_1(y) dy + \int_0^{r} \frac{v'(r;y)}{v(r;r)} \tilde{g}_2(y) dy \\
        &\geq \frac{c'_2(r)-c'_1(r)}{v(r;r)} - \max_{y \leq r} \frac{v'(r;y)}{v(r;r)} \tilde{G}_1(r) dy + \min_{y \leq r} \frac{v'(r;y)}{v(r;r)} \tilde{G}_2(r) \\
        &= \frac{c'_2(r)-c'_1(r) - \left[ \max_{y \leq r} v'(r;y) - \min_{y \leq r} v'(r;y) \right] \tilde{G}_1(r)}{v(r;r)} > 0
    \end{align*}

    If we assume $\partial v(r;y)/\partial r \partial y \geq 0$ a.e.
    \begin{align*}
        \tilde{g}_1(r)-\tilde{g}_2(r) &= \frac{c'_2(r)-c'_1(r)}{v(r;r)} - \int_0^{r} \frac{v'(r;y)}{v(r;r)} [\tilde{g}_1(y)-\tilde{g}_2(y)] dy \\
        &\geq \frac{c'_2(r)-c'_1(r)}{v(r;r)} + \frac{1}{{v(r;r)}} \int_0^r \frac{\partial v(r;y)}{\partial s \partial y} \left[ \tilde{G}_1(y)-\tilde{G}_2(y) \right] dy > 0
    \end{align*}
    where we apply integration by parts in the second line.
\end{proof}

\paragraph{Proof that WoA with costly preparation approximates WoA}

\begin{proof}
    A direct application of \eqref{eq:ieq} yields the following differential equation:
    \[
        \tilde{g}^{\epsilon}_{-i}(s) = \frac{1}{\ell_i(s) - f_i(s)} \left( \ell'_i(s) \left[ 1 - \tilde{G}^{\epsilon}_{-i}(s) \right] + \epsilon'(s) \right)
    \]
    Because this is a continuous linear mapping, we can take the limit as $\epsilon'(s)$ approaches zero. This simplifies to the same differential equation used to describe the equilibrium of the WoA (e.g. in \cite{hendricks1988war}):
    \[
        \frac{\tilde{g}_{-i}(s)}{1 - \tilde{G}_{-i}(s)} = \frac{\ell'_i(s)}{\ell_i(s) - f_i(s)}.
    \]
\end{proof}

\paragraph{Proof of Theorem \ref{thm:symmetric-nplayer}}

\begin{proof}
    Fix a symmetric auction with $n$ identical players and $m<n$ identical prizes. The expected payoff of a Player who bids $s$ is
    \[
        \int_0^s v(s;y) d\hat{G}(y) - c(s),
    \]
    $\hat{G}$ is the $n-m$ order statistic of a sample of $n-1$ draws from the equilibrium distribution, $G$. By standard arguments similar to the ones used in Lemma \ref{prop:interval}, any symmetric equilibrium will be in mixed strategies, with all players randomizing continuously on an interval $[0,\overline{s}]$ for some $\overline{s}>0$ and expected payoffs will be zero for all participants.
    Therefore, the following condition holds for all $s\in[0,\overline{s}]$:
    \[
        \int_0^s v(s;y) d\hat{G}(y) = c(s).
    \]
    As in Lemmas \ref{prop:interval} and \ref{prop:invert}, the equilibrium condition can be rewritten as
    \begin{equation}
        \hat{g}_{-i}(s) = \frac{c'_{i}(s)}{v_{i}(s; s)} - \int_0^s \frac{v'_{i}(s; y)}{v_{i}(s; s)} \hat{g}_{-i}(y) dy. \tag{\ref{eq:ieq}}
    \end{equation}
    which has a unique solution. Because this condition is the same as in the two-player case, we know that $\hat{G}$ is the two-player equilibrium. Because it is the $n-m$ order statistic of a sample of $n-1$ draws from $G$, we can write it as
    \begin{equation}
        \hat{G}(s) = \sum_{j=n-m}^{n-1} \binom{n-1}{j} [ G(s) ]^{j} [ 1 - G(s) ]^{n-j-1}. \tag{\ref{eq:orderstat}}
    \end{equation}
    All that remains is to show $G$ can be recovered from $\hat{G}(s)$. To see that this is the case, we rewrite \eqref{eq:orderstat} as $\hat{G}(s) = E[G(s)]$ where
    \[
        E[p] = \sum_{j=n-m}^{n-1} \binom{n-1}{j} p^{j} ( 1 - p )^{n-j-1}.
    \]
    That is, $E[p]$ is equal to the survival function of the binomial distribution evaluated at $n-m$. This function is known to be strictly increasing in $p$ for $p \in [0,1]$. Therefore, $E$ is invertible.
\end{proof}

\subsection{Optimal Contest Design}\label{appendixopt}

In this section, we consider how a designer should bias a contest to increase the scores. Several papers have analyzed this problem of assigning prizes to maximize total scores, or the average score of the winner. For example, \cite{mealem2014equity} consider prize redistribution in a two-player all-pay auction with fixed values and symmetric costs. They show equalizing the prize values maximizes the total scores and that the this contest yields weakly more total score than any similar Tullock-type lottery contest. \cite{che2003optimal} investigate the optimal design of contests for innovation procurement, and find that the procurer might want to limit the maximum prize available to the most efficient firms -- effectively eliminating any positive rents -- in order to increase their own expected maximum surplus. 
The problem of optimal contest design in all-pay auctions with spillovers has not been previously analyzed. 

This is relevant because principals are constrained in the prizes that they can offer. Many of the tools that principals use to make prizes have spillovers. For example, if an employer chooses to construct a compensation package using a cash bonus and stock options, then the inclusion of the stock options will generate spillovers. This section analyzes the optimal prize choice when prizes can be constructed from multiple instruments.

Let $\Lambda_i \subset \mathbb{R}^{\tilde{S}_i}$ denote the set of prize functions available to the designer for player $i$, and let $V:\prod_{i\in I}\tilde{S}_i \times \prod_{i\in I}\Lambda_i \to \mathbb{R}$ denote the designer's payoff function, i.e., given the pair of scores $\mathbf{s}:=(s_1,s_2)$ and the pair of value functions $\mathbf{v}=(v_1(\cdot;\cdot),v_2(\cdot;\cdot))$, $V(\mathbf{s},\mathbf{v})$ denotes the designer's derived net benefit from the contest.

We make the following (mild) assumptions:

\setcounter{assumption}{0}
\begin{assumption}[Completeness, D\ref{cond:D1}]\label{cond:D1}
For each $i\in I$, set of prizes $\Lambda_i$, is convex and its closure contains an element with $v_i(\cdot;\cdot) \equiv 0$.
\end{assumption}

\begin{assumption}[Productive scores, D\ref{cond:D2}]\label{cond:D2}
For each $i\in I$ and $\mathbf{v}\in \prod_{i\in I}\Lambda_i $, the designer's objective function  $ V(\mathbf{s},\mathbf{v})$ is strictly increasing in $s_{i}$.
\end{assumption}

\begin{assumption}[Costly prizes, D\ref{cond:D3}]\label{cond:D3}
For each $i\in I$, $\mathbf{s} \in \prod_{i\in I}\tilde{S}_i$ and $v_{-i}\in \Lambda_{-i}$, $ V(\mathbf{s},\mathbf{v})$ is decreasing in $v_{i}$\footnote{That is, if ${v}_i,\hat{{v}_i}\in \Lambda_i$ are such that $v_i(s;y) \leq \hat{v}_i(s;y)$ for all $(s,y) \in \tilde{S}_i\times\tilde{S}_{-i}$, then $V(\mathbf{s},(v_i,v_{-i}))\geq V(\mathbf{s},(\hat{v_i},v_{-i}))$.}. 
\end{assumption}

The primary complication with the construction in this paper is the mass point is difficult to compute. Fortunately, if the mechanism designer can discriminate between the two players, an optimal mechanism will have no atoms in many specifications. This is formalized in the following proposition.

\begin{prop}\label{prop:mecn}
Assume a two-player contest where a fully informed principal with payoff function $V$ chooses the prize $v_i\in\Lambda_i$ for each $i\in I$. Assume that $\Lambda_i$ and $V$ satisfy assumptions D\ref{cond:D1} to D\ref{cond:D3}, and that for all $i$ and all $v_i \in \Lambda_i$, assumptions A\ref{cond:smo} to A\ref{cond:pos} hold. Then, no contestant in equilibrium can have a positive payoff. Equivalently, no player will have a point-mass as part of their strategy. 
\end{prop}

Proposition \ref{prop:mecn} implies that there will be no strictly dominant player in any discriminating contest design problem where the principal benefits from the efforts of participants and pays for prizes. This proposition comes from the fact that the equilibrium strategy of the dominant player is locally invariant to changes in her prize value. Intuitively, for any contest with a strictly dominant player, there exists a more competitive contest where their prize is reduced and scores are larger.

\begin{proof}
Take an optimal choice of $\mathbf{v}:=(v_i)_{i\in I} \in \prod_{i\in I}\Lambda_i$. Suppose, by contradiction, that player $i$ has a strictly positive payoff. Her strategy is defined by
\[
    \tilde{g}_{i}(s) = \frac{c'_{-i}(s)}{v_{-i}(s; s)} - \int_0^s \frac{v'_{-i}(s; y)}{v_{-i}(s; s)} \tilde{g}_{i}(y) dy,
\]
which does not depend on $v_i$. Because player $-i$ has an atom, we know that $\tilde{G}_i(\bar{s}) - \tilde{G}_{-i}(\bar{s}) > 0$. Therefore, there exists a $\gamma \in (0,1)$ such that $\tilde{G}_i(\bar{s}) = \tilde{G}_{-i}(\bar{s})/\gamma$.

Then, the principal could offer $(\gamma v_i, v_{-i})$ without changing the equilibrium strategy of player $i$. By the costly prizes Assumption D\ref{cond:D3}, this   is weakly preferable given a fixed distribution of $s_{-i}$.

By construction, player 2's new equilibrium strategy is $\tilde{G}_{-i}(\bar{s})/\gamma$. This first-order stochastically dominates player $-i$'s original strategy. In fact, it is the same distribution, but with the mass point removed. The productive scores assumption implies that this mechanism is strictly preferred.
\end{proof}

Proposition \ref{prop:mecn} demonstrates that the expected welfare of all agents is zero in a large class of contest design problems\footnote{Which is not to say that there are no  settings where it would not apply to. For example, the designer could wish to maximize the agents' expected welfare. In this case, the principal's objective function would violate costly prizes. It would usually also violate productive scores.}. It also suggests the optimality, from a design perspective, of handicapping the most efficient players (as in, the players with lower costs and lower marginal costs) The idea is very much analogous to the conclusion in \cite{che2003optimal}, for example: handicapping the player that has the technological upper hand causes the less efficient player to become more aggressive, and to choose higher scores than they would otherwise.

\subsection{Removing assumptions}\label{appendix:assumptions}

An auction which satisfies all assumptions A\ref{cond:smo}--\ref{cond:pos} but one may have equilibria which fail to meet our characterization. An auction which fails to satisfy any of A\ref{cond:mon}--\ref{cond:pos} can have multiple equilibria. 

\paragraph{Assumption 1 (Smoothness)} We assume continuous differentiability of $v_i$ and $c_i$. Continuity is not sufficient to ensure that the equilibrium has interval support. For example, consider the case where the prize is fixed at $v_i=1$ and the costs are given by the density function of some distribution which uniformly assigns probability one to a dense subset of $[0,1]$ with Lebesgue measure zero.\footnote{For example a uniform distribution over the countable union of cantor sets shifted by each of the rationals modulo one.} This cost function is continuous because the distribution assigns uniform weight to infinitely many points. It is also strictly increasing because the support is dense. However, it is not absolutely continuous. Then, the aforementioned distribution is an equilibrium, which has support only on a set of measure zero.

\paragraph{Assumption 2 (Monotonicity)} The case where $v'_i(s_i;y) > c'_i(s_i)$ for some $s_i$ is considered in \citet{siegel2014contests} without spillovers. In this case, the equilibrium distribution has gaps and is thus not an interval. In the presence of spillovers, non-monotonicity may generate pure-strategy equilibria or result in non-uniqueness. For example, consider the symmetric game where 
\[
    v_1(s;y)=v_2(s;y)=v(s;y) =
    \begin{cases}
    1 + s - y &\text{if } s \leq 1 \\
    (3-s)s - y  &\text{if } s >1
    \end{cases}
\]
and $c_1(s)=c_2(s)=c(s) = s$. Note that this prize value satisfies all assumptions except for monotonicity, which is violated on $[0,1]$. There are two asymmetric pure strategy equilibria where one player bids 0 and the other player bids 1.\footnote{This game also has a symmetric mixed-strategy equilibrium.}

\paragraph{Assumption 3 (Interiority)} Consider the symmetric all-pay auction with spillovers where $v_1(s;y)=v_2(s;y)=v(s;y) = 2 \sqrt{y}$ and $c_1(s)=c_2(s)=c(s) = s$. Then, there is a pure strategy equilibrium where both players play zero. Moreover, there is also a mixed strategy equilibrium at
\[
    G_1^\star(x) = G_2^\star(x) =
    \begin{cases}
    \sqrt{x} &\text{if } x \in [0,1] \\
    0        &\text{if } x < 0 \\
    1        &\text{if } x > 1,
    \end{cases}
\]
i.e. equilibria are no longer unique.

In the other case where costs are no higher than the prize value in the limit (that is, $\lim_{s_i\to \infty}\sup_{y\in\tilde{S}_{-i}}v_i(s_i;y)\geq \lim_{s_i\to \infty}c_i(s_i)$), players might find it profitable to submit unbounbed bids. This can result in non-existence.

\paragraph{Assumption 4 (Discontinuity at ties)} Consider the symmetric all-pay auction with spillovers where $v_1(s;y)=v_2(s;y)=v(s;y) = \mathbf{1}_{y\leq 1} 4 \sqrt{1-y}$ and $c(s) = s$. Then, there is a symmetric equilibrium where 
\[
    G_1^\star(x) = G_2^\star(x) =
    \begin{cases}
    \frac{1 - \sqrt{1-x}}{2} &\text{if } x \in [0,1) \\
    0        &\text{if } x < 0 \\
    1        &\text{if } x \geq 1
    \end{cases}
\]
such that both players have an atom at one, the point where the prize is worth zero in the event of a tie. The usual argument that the two players cannot have atoms at the same point fails here because a small increase in either player's bid does not increase the probability of winning a prize \textit{of positive value}. There are also two asymmetric equilibria in this game of the form
\[
    G_i^\star(x) =
    \begin{cases}
    \frac{1 - \sqrt{1-x}}{2} &\text{if } x \in [0,1) \\
    0        &\text{if } x < 0 \\
    1        &\text{if } x \geq 1
    \end{cases}
    \qquad
    G_{-i}^\star(x) =
    \begin{cases}
    1 - \frac{\sqrt{1-x}}{2} &\text{if } x \in [0,1] \\
    0        &\text{if } x < 0 \\
    1        &\text{if } x > 1.
    \end{cases}
\]
That is, one player has an atom at zero while the other has an atom at 1. Any convex combination of the symmetric equilibrium and the above is also an equilibrium.

\subsection{Numerical Approximation}\label{appendixnum}

\paragraph{Iteration method}
It is possible to approximate the solution by iterating numerically on this sequence:
\[
    \tilde{g}_{n+1}(s) = \frac{1}{v(s; s)} \left( c'(s) - \int_0^s v'(s; y) \tilde{g}_n(y) dy \right)
\]
starting from $\tilde{g}_0 = 0$ to find the true $\tilde{g}$. There is a much simpler and faster way.

\paragraph{Matrix method (1)}
Consider our original equation
\[
    \int_0^s v_{-i}(s; y) \tilde{g}_i(y) dy = c(s)
\]
and consider this $3 \times 3$ discrete approximation of this problem for $s \in [0,1]$
\[
    \frac{1}{3}
    \underbrace{\left[
\begin{matrix}
    v_{-i}(\frac{1}{3},\frac{1}{3}) & 0                              & 0   \\
    v_{-i}(\frac{2}{3},\frac{1}{3}) & v_{-i}(\frac{2}{3},\frac{2}{3}) & 0    \\
    v_{-i}(1,\frac{1}{3})            & v_{-i}(1,\frac{2}{3})            & v_{-i}(1,1)
\end{matrix}    
    \right]}_{\mathbf{V}}
\cdot
    \underbrace{\left[
\begin{matrix}
    \tilde{g}_i(\frac{1}{3}) \\
    \tilde{g}_i(\frac{2}{3})  \\
    \tilde{g}_i(1)            
\end{matrix}    
    \right]}_{\mathbf{g}}
\approx
    \underbrace{\left[
\begin{matrix}
    c_{-i}(\frac{1}{3}) \\
    c_{-i}(\frac{2}{3})  \\
    c_{-i}(1)            
\end{matrix}    
    \right]}_{\mathbf{c}}
\]

So, we can approximate $\tilde{g}_i(s)$ with
\[
    \mathbf{g} = 3 \mathbf{V}^{-1} \mathbf{c}
\]

\paragraph{Matrix method (2)}
To get a good estimate, we do the same thing with an $N \times N$ grid for $N$ large on some interval $[0,T]$.\footnote{The Python package \href{https://pypi.org/project/allpy/}{\texttt{allpy}} implements the approximation algorithm in this section and computes mixed-strategy equilibria of all-pay auctions with spillovers}
\[
    \left[
\begin{matrix}
    \tilde{g}_i(\frac{1}{N}) \\
    \tilde{g}_i(\frac{2}{N})  \\
    \vdots \\
    \tilde{g}_i(T)            
\end{matrix}
    \right]
    \approx
    N
    \left[
\begin{matrix}
    v_{-i}(\frac{1}{N},\frac{1}{N}) & 0         & \cdots & 0   \\
    v_{-i}(\frac{2}{N},\frac{1}{N}) & v_{-i}(\frac{2}{N},\frac{2}{N}) &  \cdots & 0    \\
     &  & \ddots & \\
    v_{-i}(T,\frac{1}{N})            & v_{-i}(T,\frac{2}{N})  &    \cdots     & v_{-i}(T; T)
\end{matrix}    
    \right]^{\textstyle-1}
\cdot
    \left[
\begin{matrix}
    c_{-i}(\frac{1}{N}) \\
    c_{-i}(\frac{2}{N})  \\
    \vdots \\
    c_{-i}(T)            
\end{matrix}    
    \right]
\]

\paragraph{Getting the actual strategies}
Once you get $(\tilde{g}_1,\tilde{g}_2)$ you just have to:\footnote{Sample Python code provided below each item.}
\begin{enumerate}
    \item take the cumulative sum and divide by $N$ to get $(\tilde{G}_1,\tilde{G}_2)$
\begin{verbatim}
G1, G2 = cumsum(g1)/N, cumsum(g2)/N
\end{verbatim}
    \item truncate both distributions at the last value where both are $\leq 1$
\begin{verbatim}
G1, G2 = G1[G1 <= 1 & G2 <= 1], G2[G1 <= 1 & G2 <= 1]
\end{verbatim}
    \item add to each CDF vector so that both end with 1 (add the atom)
\begin{verbatim}
G1, G2 = (G1 - G1[-1] + 1), (G2 - G2[-1] + 1)
\end{verbatim}
\end{enumerate}

\end{document}